\documentclass[twocolumn,published,twocolappendix]{aastex631} 

\usepackage[T1]{fontenc}
\usepackage{savesym}
\savesymbol{tablenum}
\usepackage{siunitx}
\restoresymbol{SIX}{tablenum}
\usepackage{multirow}

\shorttitle{Planetesimal formation under SI in HD~163296}
\shortauthors{F.~Zagaria et al.}

\begin{document}

\title{Observing planetesimal formation under streaming instability in the rings of HD 163296}

\correspondingauthor{F. Zagaria}\email{fz258@cam.ac.uk}
\author[0000-0001-6417-7380]{F. Zagaria}
\affiliation{Institute of Astronomy, University of Cambridge, Madingley Road, Cambridge CB3 0HA, UK}
\author[0000-0003-4288-0248]{C. J. Clarke}
\affiliation{Institute of Astronomy, University of Cambridge, Madingley Road, Cambridge CB3 0HA, UK}
\author[0000-0002-0364-937X]{R. A. Booth}
\affiliation{School of Physics and Astronomy, University of Leeds, Leeds, LS2 9JT, UK}
\author[0000-0003-4689-2684]{S. Facchini}
\affiliation{Dipartimento di Fisica, Università degli Studi di Milano, Via Giovanni Celoria 16, 20133 Milano, Italy}
\author[0000-0003-4853-5736]{G. P. Rosotti}
\affiliation{Dipartimento di Fisica, Università degli Studi di Milano, Via Giovanni Celoria 16, 20133 Milano, Italy}



\begin{abstract}
We introduce a new technique to determine the gas turbulence and surface density in bright disc rings, under the assumption that dust growth is limited by turbulent fragmentation at the ring centre. We benchmark this prescription in HD~163296, showing that our measurements are consistent with available turbulence upper limits and agree with independent estimates of the gas surface density within a factor of two. We combine our results with literature measurements of the dust surface density and grain size to determine the dust-to-gas ratio and Stokes number in the 67$\,$au and 100$\,$au rings. Our estimates suggest that particle clumping is taking place under the effect of streaming instability (SI) in the 100$\,$au ring. Even though in the presence of external isotropic turbulence this process might~be hindered, we provide evidence that turbulence is non-isotropic in both rings and likely originating from mechanisms (such as ambipolar diffusion) that could ease particle clumping under SI. Finally, we determine the mass accretion rate under the assumption that the disc is in steady state and turbulence regulates angular momentum transport. Our results are in tension with spectroscopic measurements and suggest that other mechanisms might be responsible for accretion, in qualitative agreement with the detection of a magneto-centrifugal wind in this system. Applying our method to larger samples can be used to statistically assess if SI is a viable mechanism to form planetesimals in bright rings. 
\end{abstract}




\section{Introduction}\label{sec:introduction}
Planets form in gas- and dust-rich discs orbiting young stars. According to the core-accretion model, this process takes place sequentially. At first, gentle collisions among $\mu$m-sized grains are expected to promote dust coagulation into mm- to cm-sized pebbles (e.g., \citealt{Brauer2008,Birnstiel2010}), in broad agreement with laboratory experiments (see \citealt{Testi2014,Birnstiel2016}) and (sub-)mm continuum observations (e.g., \citealt{Tazzari2016,Tazzari2021,Carrasco-Gonzalez2019,Macias2021,Sierra2021,Guidi2022}). However, larger grains are subject to larger relative particle velocities, that can halt further dust coagulation \citep{Brauer2008,Birnstiel2010} because of non-adhesive (bouncing, \citealt{Zsom2010}) or destructive (turbulent fragmentation, \citealt{Ormel&Cuzzi2007}) collisions and radial drift \citep{Weidenschilling1977,Nakagawa1986}. How pebbles overcome these growth barriers to form km-sized bodies, the so-called planetesimals, is still a matter of debate.

The streaming instability (SI, \citealt{Youdin&Goodman2005}, see also \citealt{Lesur_ppvii,Simon2022}) is a promising solution to this conundrum. SI is a two-fluid resonant drag instability that arises from the differential rotation of gas and dust in the disc mid-plane (e.g., \citealt{Squire&Hopkins2020}). In a cylindrical shear flow with radially decreasing pressure gradient, gas rotates with sub-Keplerian velocity. Dust, instead, is not pressure-supported and orbits with Keplerian speed. Because of this azimuthal velocity difference, the gas drag reduces the angular momentum of solids, that decouple from the background gas and drift radially inwards. However, when dust backreaction on gas is strong enough, it increases the gas azimuthal velocity, reducing the dust drift efficiency \citep{Johansen&Youdin2007} and favouring the pile-up of solids \citep{Youdin&Johansen2007}. In the presence of dust overdensities, this mechanism can rapidly lead to strong particle concentration, creating narrow dust filaments dense enough to favour the collapse of self-gravitating particle clumps that will eventually form planetesimals \citep{Johansen2007,Johansen2009}. 

Among the other planetesimal formation mechanisms proposed in the last years (e.g., hierarchical coagulation models, such as porous growth or growth by mass transfer, see \citealt{Johansen_ppvi,Blum2018,Drazkowska_ppvii} and references therein), SI appears to be the most promising one, because it is consistent with a number of Solar System observations, such as the structure of comets \citep{Blum2017}, the prograde rotation of trans-Neptunian objects \citep{Nesvorny2019}, the formation of contact binaries in the cold classical Kuiper Belt (e.g., Arrokoth, \citealt{McKinnon2020}) and its absolute magnitude distribution \citep{Kavelaars2021}.

Shearing box simulations of non-linear dust-gas interactions in vertically stratified discs revealed that particle clumping under SI is governed by three main parameters: (i) the local dust-to-gas surface density ratio,
\begin{equation}\label{eq:1}
    Z=\Sigma_{\rm dust}/\Sigma_{\rm gas},
\end{equation}
where $\Sigma_{\rm gas}$ and $\Sigma_{\rm dust}$ are the gas and dust surface densities, that determines the efficiency of dust backreaction; (ii) the particle Stokes number, ${\rm St}$, that describes the degree of coupling between gas and dust; (iii) the pressure support,
\begin{equation}\label{eq:2}
    \Pi=\frac{\Delta v}{c_{\rm s}}=-\frac{1}{2}\frac{c_{\rm s}}{v_{\rm K}}\frac{d\ln P}{d\ln R},
\end{equation}
that determines the relative azimuthal velocity between gas and dust. Here $\Delta v=v_{\rm K}-v_\varphi$ is the gas azimuthal velocity ($v_\varphi$) deviation from Keplerian rotation ($v_{\rm K}$), $c_{\rm s}$ is the locally isothermal sound speed, $P$ is the gas pressure and $R$ is the disc radial coordinate. Recent studies \citep{Carrera2015,Yang2017,Li&Youdin2021} showed that, at a fixed value of $\Pi$, particle clumping under SI is favoured when ${\rm St}\approx0.1$, 
but it requires progressively larger dust-to-gas ratios for smaller values of the Stokes number. 
Steeper pressure gradients are also expected to hinder particle clumping \citep{Johansen2007,Bai&Stone2010_pressure,Sekiya&Onishi2018}.

Evidence that discs frequently display substructures in dust continuum emission \citep{Long2018,Andrews2018,Andrews2020} suggests that bright rings could be sweet spots for planetesimal formation under SI. Indeed, gaps are expected to halt radial drift, piling solids up in bright rings, where the dust-to-gas ratio might be locally strongly enhanced, favouring SI-driven particle clumping. This popular hypothesis is~supported by the evidence that (at least some) bright rings are pressure traps \citep{Dullemond2018,Rosotti2020,Izquierdo2023} and was successfully invoked to explain the optical depth of DSHARP rings \citep{Stammler2019}. 

Even though the conditions for particle clumping in pressure bumps were studied using vertically stratified shearing box simulations in a number of physical settings \citep{Carrera2021,Carrera2022,Carrera&Simon2022,Xu&Bai2022b}, no direct method to observationally assess if SI-driven planetesimal formation is underway in bright disc rings has been identified yet. Recently, \citet{Scardoni2021} proposed that particle clumping under SI might have observable effects, reducing the disc optical depth and affecting the (sub-)mm spectral index depending on the size and opacity of the particles forming clumps. They also showed that the~presence of SI-driven dust accumulations is consistent with the \textit{optical} properties of Lupus discs. A more direct way to assess if SI is a robust mechanism to form planetesimals in bright rings, would be to compare their \textit{physical} properties (i.e., $Z$, ${\rm St}$) with the available thresholds for particle clumping \citep{Carrera2015,Yang2017,Li&Youdin2021}. However, measurements of the gas surface density, which are precluded in most cases, are needed to estimate such quantities. 

In this Letter, we present a new analysis technique that combines information on dust temperature, density and grain size from multi-frequency dust continuum observations with knowledge of the dust-to-gas coupling to measure the dust-to-gas ratio and Stokes number in the 67$\,$au and 100$\,$au ring of HD~163296, under the assumption that dust growth is fragmentation limited. In \autoref{sec:methods} we introduce our method and justify its application to HD~163296. In \autoref{sec:results} and \ref{sec:discussion} we present and discuss our results and, finally, in \autoref{sec:conclusion} we draw our conclusions and consider future prospects.

\begin{table*}[t!]
    \centering
    \begin{tabular}{|c|c|c|c|c|c|c|c|c|}
    \hline
            & $\alpha_r/{\rm St}$ & $T$ (K) & $\log(a_{\rm frag}/{\rm cm})$ & $\alpha_{\rm turb}\times10^4$ & $\Sigma_{\rm gas}$ (g$\,$cm$^{-2}$) & $\Sigma_{\rm dust}$ (g$\,$cm$^{-2}$) & $Z\times10^2$ & ${\rm St}_{\rm frag}\times10^3$\\
    \hline
    \hline
    R67     & $0.23\pm0.03$              & $21.94\pm0.10$ & $-1.88\pm0.11$ & $6.09\pm0.40$ & $16.11\pm4.32$ & $0.267\pm0.002$ & $1.66\pm0.45$ & $2.65\pm0.17$\\
    R100    & $0.04\pm0.01$              & $12.83\pm0.08$ & $-1.90\pm0.13$ & $3.32\pm0.42$ &  $4.98\pm1.64$ & $0.454\pm0.012$ & $9.13\pm3.02$ & $8.31\pm1.04$\\
    \hline
    \end{tabular}
    \caption{Ring parameters constrained from multi-frequency dust and gas kinematics observations. (1) Ring ID. (2) Dust-to-gas coupling (from \citealt{Rosotti2020}). (3)--(4) Dust temperature and maximum grain size (from \citealt{Guidi2022}). (5) Turbulent diffusivity. (6) Gas surface density. (7) Dust surface density (from \citealt{Guidi2022}). (8) Dust-to-gas ratio. (9) Stokes number.}
    \label{tab:1}
\end{table*}

\section{Methods}\label{sec:methods}
In the hypothesis that particle fragmentation due to turbulent relative motions sets the maximum grain size at the ring centre, we can write \citep{Birnstiel2012}
\begin{equation}\label{eq:3}
    a_{\rm max}\equiv a_{\rm frag} = 0.37\frac{2}{3\pi}\frac{\Sigma_{\rm gas}}{\rho_{\rm s}\alpha_{\rm turb}}\left(\frac{u_{\rm frag}}{c_{\rm s}}\right)^2,
\end{equation}
where $\rho_{\rm s}$ is the dust material density, $u_{\rm frag}$ is the velocity threshold for dust fragmentation and $\alpha_{\rm turb}$ regulates the turbulent collision velocity of dust particles \citep{Ormel&Cuzzi2007}. The Stokes number of a compact particles with size $a_{\rm frag}$ near the disc mid-plane can be written in the Epstein regime as
\begin{equation}\label{eq:4}
    {\rm St}_{\rm frag} = \frac{\pi}{2}\frac{a_{\rm frag}\rho_{\rm s}}{\Sigma_{\rm gas}}\rightarrow
    \frac{{\rm St}_{\rm frag}}{\alpha_{\rm turb}} = \frac{\pi}{2}\frac{a_{\rm frag}\rho_{\rm s}}{\Sigma_{\rm gas}\alpha_{\rm turb}}.
\end{equation}

$a_{\rm frag}$ in \autoref{eq:3} and $\alpha_{\rm turb}/{\rm St}_{\rm frag}$ in \autoref{eq:4} scale with the ratio and product of $\Sigma_{\rm gas}$ and $\alpha_{\rm turb}$, respectively, thus combining these expressions allows us to disentangle the two and write an expression for turbulent diffusivity
\begin{equation}\label{eq:5}
    \alpha_{\rm turb} = 0.35\frac{u_{\rm frag}}{c_{\rm s}}\left(\frac{\alpha_{\rm turb}}{{\rm St}_{\rm frag}}\right)^{1/2},
\end{equation}
and gas surface density
\begin{equation}\label{eq:6}
    \Sigma_{\rm gas} = 2.85\frac{\pi}{2}\rho_{\rm s}a_{\rm frag}\frac{c_{\rm s}}{u_{\rm frag}}\left(\frac{\alpha_{\rm turb}}{{\rm St}_{\rm frag}}\right)^{1/2}.
\end{equation}

These expressions show that $\Sigma_{\rm gas}$ and $\alpha_{\rm turb}$ can be determined from two sets of parameters. (i) The maximum grain size ($a_{\rm max}$) and the dust mid-plane temperature ($c_{\rm s}\propto T^{1/2}$), that can be measured using high angular resolution multi-frequency dust continuum observations. (ii) The gas-to-dust coupling ($\alpha_{\rm turb}/{\rm St}_{\rm frag}$, under the assumption that dust radial diffusivity, $\alpha_r$, is regulated by gas turbulence, i.e. $\alpha_r=\alpha_{\rm turb}$). This can be inferred from the dust and gas ring widths, the latter of which is derived from the gradient of the azimuthal velocity deviation from Keplerian rotation \citep{Dullemond2018,Rosotti2020}, measured from the rotation curve of bright emission lines observed at high spectral resolution.

\section{Results}\label{sec:results}
As a proof of concept, we computed $\Sigma_{\rm gas}$ and $\alpha_{\rm turb}$ in the 67$\,$au and 100$\,$au rings of HD~163296 (e.g., \citealt{Isella2018}), a 6~Myr-old, 1.95~$M_\odot$ young stellar object \citep{Wichittanakom2020}, located 100.96~pc away \citep{Gaia_EDR3}, and among the best studied in the (sub-)mm. This is the \textit{only} source with state-of-the-art CO emission line data (e.g., \citealt{Teague2018,Teague2021,Izquierdo2022,Izquierdo2023}) where dust properties were measured using multi-frequency ALMA and VLA continuum observations \citep{Guidi2022}. 

We used the dust-to-gas coupling estimated by \citet{Rosotti2020}\footnote{The slope of the gas azimuthal velocity deviation from Keplerian rotation induced by the ring ($\delta v_\varphi=v_\varphi/v_{\rm K}-1$) measured by \citet{Rosotti2020} using DSHARP data \citep{Andrews2018} are broadly consistent with those we estimated from the rotation curve of \citet{Izquierdo2023} using higher spectral resolution MAPS data \citep{Oberg2021}.} in combination with the temperature and grain size determined by \citet{Guidi2022}. We adopted a dust material density $\rho_{\rm s}=\SI{2.08}{\gram\per\cubic\cm}$, since \citet{Guidi2022} assumed dust to be composed by 60\% amorphous silicates (Mg$_{0.7}$Fe$_{0.3}$SiO$_3$, \citealt{Dorschner1995}), 15\% amorphous carbon \citep{Zubko1996} and 25\% porosity by volume, and a fragmentation velocity $u_{\rm frag}=\SI{1}{\m\per\s}$, motivated by recent laboratory experiments \citep{Gundlach2018,Musiolik&Wurm2019}. Columns (1)-(6) of \autoref{tab:1} summarise these parameters and our newly inferred $\Sigma_{\rm gas}$ and $\alpha_{\rm turb}$. 

Independent literature measurements of these quantities can be used to benchmark our procedure. \citet{Flaherty2017} estimated the turbulent broadening of CO isotopologue and DCO$^+$ emission lines in the outer disc of HD~163296, showing that $\alpha_{\rm turb}<3\times10^{-3}$, consistent with our results. \citet{Booth2019} forward modelled $^{13}$C$^{17}$O emission in HD~163296 to constrain the total gas disc mass. Their fiducial gas surface density 
in the 67$\,$au and 100$\,$au rings is reported in Column (2) of \autoref{tab:2} and shows a remarkable agreement with our estimates, falling within $1\sigma$ in R67 and $3\sigma$ in R100 (about a factor of two off).

\begin{table}[t!]
    \centering
    \begin{tabular}{|c|c|c|c|}
    \hline
            & $\Sigma_{\rm gas}$ (g$\,$cm$^{-2}$) & $Z\times10^2$ & ${\rm St}_{\rm frag}\times10^3$\\
    \hline
    \hline
    R67     & 19.20 & $1.39\pm0.01$ & $2.22\pm0.58$ \\
    R100    & 10.33 & $4.40\pm0.12$ & $4.00\pm1.22$ \\
    \hline
    \end{tabular}
    \caption{Same as in \autoref{tab:1}, using the CO-based gas surface density of \citet{Booth2019}. (1) Ring ID. (2) Gas surface density. (3) Dust-to-gas ratio. (4) Stokes number.}
    \label{tab:2}
\end{table}

\begin{figure}[t!]
    \centering
    \includegraphics[width=\columnwidth]{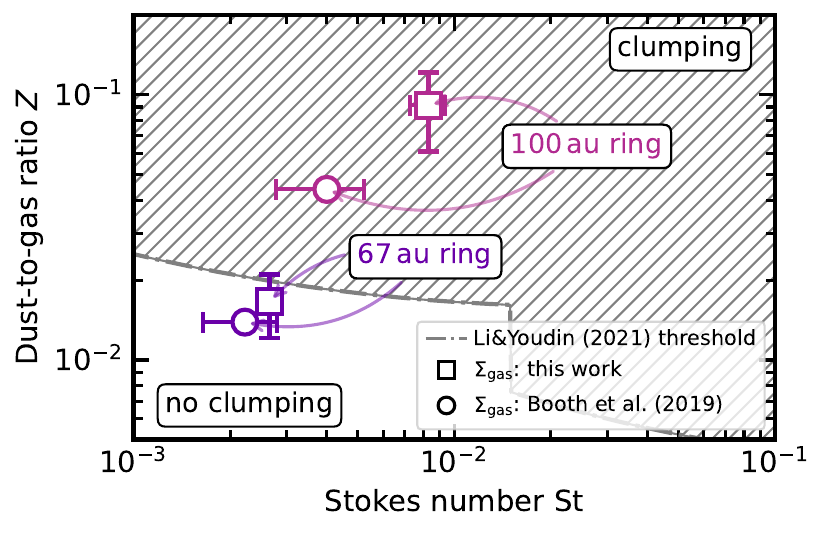}
    \caption{Dust-to-gas ratio and Stokes number of the 67$\,$au (violet) and 100$\,$au (purple) rings of HD~163296 plotted against the threshold for particle clumping in laminar discs ($\alpha_{\rm turb}=0$) of \citet{Li&Youdin2021}. Squares display the values determined using the gas surface density from \autoref{tab:1}, while circles show those computed from the gas mass measured by \citet[][see \autoref{tab:2}]{Booth2019}. The 100$\,$au ring is prone to particle clumping under SI.}
    \label{fig:1a}
\end{figure}

We can use the gas surface density in \autoref{tab:1} together with the grain size and dust surface density measured by \citet{Guidi2022} to determine $Z$ and ${\rm St}$ from \autoref{eq:1} and \ref{eq:4}. Our results are summarised in Columns (7)-(9) of \autoref{tab:1}, while in Columns (3)-(4) of \autoref{tab:2} we report the values of $Z$ and ${\rm St}$ computed using the gas surface density of \citet{Booth2019} as a sanity check. Are these parameters consistent with SI forming particle clumps in the rings of HD~163296? In \autoref{fig:1a} we show a comparison of the dust-to-gas ratio and Stokes number from \autoref{tab:1} (squares) and \autoref{tab:2} (dots) in R67 (violet) and R100 (purple) with the threshold for particle clumping under SI proposed by \citet{Li&Youdin2021}. The plot shows that SI-driven particle clumping is underway in the 100$\,$au ring, as predicted by \citet{Stammler2019}, both for the surface density from \autoref{eq:6} and the CO-based estimate of \citet{Booth2019}.


\section{Discussion}\label{sec:discussion}
Hereafter, we discuss how the gas radial pressure gradient, the disc turbulence and the particle size distribution can influence our results. We also comment on the assumptions we made in our derivation of the gas surface density and possible implications for disc evolution.

\subsection{Pressure support}
The threshold for particle clumping by SI we adopted in \autoref{fig:1a} was estimated with a suite of vertically stratified 2D shearing box simulations with $\Pi=0.05$ \citep{Li&Youdin2021}. However, since the 67$\,$au and 100$\,$au~rings of HD~163296 are pressure maxima \citep{Rosotti2020}, we should compare our results with the threshold for $\Pi\rightarrow0$. In fact, \citet{Bai&Stone2010_pressure} showed that in this regime particle clumping under SI requires lower values of $Z$. On the other hand, SI needs a non-null pressure gradient to operate, otherwise no azimuthal velocity difference between dust and gas can sustain particle drift into local overdensities\footnote{We stress, however, that even though SI cannot operate at the ring centre, it could still take place in the wings.}. 

\citet{Carrera2021,Carrera2022} recently performed 3D shearing box simulations of SI-driven particle clumping in traffic jams (i.e., small-amplitude pressure bumps inducing local dust accumulations, but not able to halt radial drift entirely). Surprisingly, they showed that solids with ${\rm St}=0.012$ (similar to those in \autoref{tab:1}) never produce clumps for bump amplitudes as large as 50\% \citep{Carrera&Simon2022}, even though their local dust-to-gas ratio is well above the threshold proposed by \citet{Li&Youdin2021}. Instead, pressure bumps with amplitudes larger than 50\% can lead to substantial particle clumping. In this case, however, SI is not needed to concentrate solids because the bump is strong enough to efficiently form planetesimals purely by gravitational instability \citep{Carrera&Simon2022}. 

Since \citet{Rosotti2020} showed that R67 and R100 are particle traps (corresponding to bump amplitudes larger than 70\% in the simulations of \citealt{Carrera&Simon2022}), the latter scenario seems to be closer to our case, suggesting that planetesimals could form through GI rather than SI in these rings. However, none of the previously cited simulations considered how turbulence could affect particle clumping. Our results, instead, are based on the assumption that the 67$\,$au and 100$\,$au rings of HD~163296 are in a steady-state balance between dust trapping and diffusion (see also \citealt{Rosotti2020}).

\subsection{Turbulence}
Turbulence, even at small levels, can be detrimental for particle clumping, reducing the mid-plane solid density and diffusing local dust enhancements. Although~SI itself induces radial and vertical diffusivity (e.g., \citealt{Johansen&Youdin2007,Bai&Stone2010_SI}), the radial ($\alpha_r$) and vertical ($\alpha_z$) SI diffusion coefficients are $\alpha_z\lesssim\alpha_r\lesssim10^{-6}$ for ${\rm St}\lesssim10^{-2}$ \citep{Li&Youdin2021}. Considering this value as the minimum diffusivity needed to interfere with SI, the turbulent levels we measured in the rings of HD~163296 (see \autoref{tab:1}) are high enough to affect the threshold for particle clumping in \autoref{fig:1a}.

\begin{figure}[t!]
    \centering
    \includegraphics[width=\columnwidth]{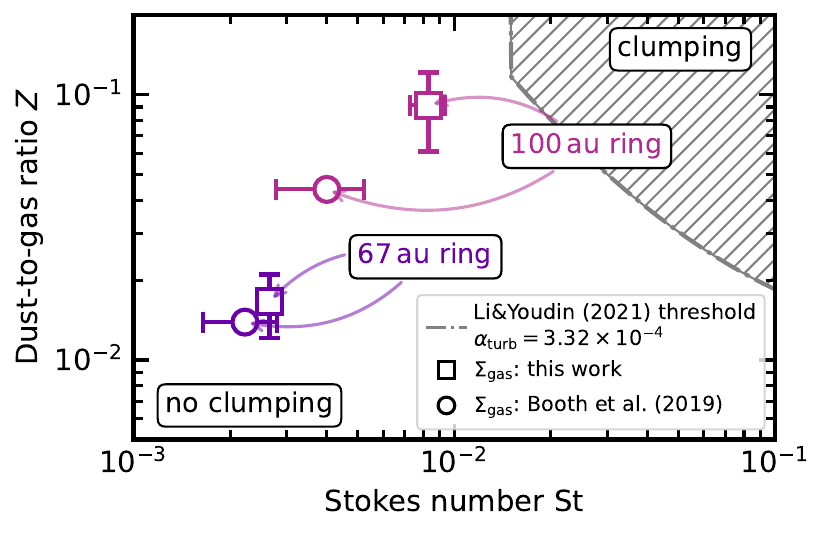}
    \caption{Same as \autoref{fig:1a}, but showing the threshold for particle clumping of \citet{Li&Youdin2021} corrected for an isotropic turbulence level of $\alpha_{\rm turb}=3.32\times10^{-4}$. Both rings are stable against SI-driven particle clumping. However, we stress that turbulence is non-isotropic in R67 and R100, which leads to less stringent conditions for particle clumping.}
    \label{fig:1b}
\end{figure}

To quantify the impact of turbulence on SI-driven particle clumping, \citet{Gole2020} performed 3D vertically stratified simulations including a forcing term in the gas momentum equation to generate an isotropic Kolmogorov-like turbulence. They proposed that turbulence primarily affects SI by increasing the thickness of the solid layer, changing the mid-plane dust-to-gas density ratio. \citet{Li&Youdin2021} modified their threshold for particle clumping, including the effect of isotropic turbulence to reproduce the results of \citet{Gole2020}. 
In \autoref{fig:1b}, this new threshold for $\alpha_{\rm turb}=\alpha_{\rm turb}(R=100\ {\rm au})$ is plotted against the dust-to-gas ratio and Stokes number measured in R67 and R100. The plot shows that isotropic turbulence halts particle clumping under SI in both rings.

However, the turbulence implementation described by \citet{Gole2020} is idealised, since it does not take into account how magnetic fields could modify the properties of turbulence and how this is affected by particle overdensities. In fact, several works have argued that a more realistic treatment of external turbulence, such as non-ideal magneto-hydrodynamic (MHD) effects (e.g., Ohmic dead zones, \citealt{Yang2018}) or the vertical shear instability (VSI, \citealt{Schafer2020}), might even be beneficial for particle clumping, seeding SI in by large scale effects, such as zonal flows or non-isotropic diffusivity. 

\citet{Xu&Bai2022a} recently performed vertically stratified 3D shearing box non-ideal MHD simulations including the effects of ambipolar diffusion (AD). They found that the spontaneous formation of zonal flows promotes SI-driven particle clumping under less stringent conditions than in the laminar case, especially for shallower pressure gradients, showing that SI-driven planetesimal formation can take place with $\Pi=0$ when AD is active. Additionally, \citet{Xu&Bai2022b} found strong dust clumping in the case of pressure maxima with amplitudes $\leq$50\%. Although similar to those of \citet{Carrera&Simon2022}, these bumps achieved a quasi-steady state balance between dust trapping and the turbulent diffusion generated by AD, as expected from our measurements (see \autoref{tab:1}). In the absence of dust feedback, non-ideal MHD simulations with AD predict that vertical diffusion is larger than radial diffusion, $\alpha_z\gg\alpha_r$, because AD increases the eddy turnover time in the vertical direction \citep{Xu&Bai2022a}. Instead, when dust backreaction is included, especially if particle clumping in pressure bumps takes place, it preferentially reduces the correlation time of vertical turbulent fluctuations, locally leading to $\alpha_z\ll\alpha_r$ \citep{Xu&Bai2022b}. 

To assess if SI-driven particle clumping in the presence of AD is a viable scenario for HD~163296, we adopted literature measurement of the dust-to-gas coupling in the vertical direction \citep{Doi&Kataoka2021,Liu2022} to determine the vertical diffusivity in the 67 and 100$\,$au rings of HD~163296 using the Stokes numbers from \autoref{tab:1}. Our results are summarised in \autoref{tab:3} and, when compared with our estimates of the radial diffusivity in \autoref{tab:1}, they suggest that turbulence is non-isotropic in both rings (see also \citealt{Doi&Kataoka2023}). 

\begin{table*}[t!]
    \centering
    \begin{tabular}{|c|c|c|c|c|}
    \hline
            & $\alpha_z/{\rm St}$ & $\alpha_z\times10^4$ & $\alpha_z/{\rm St}$ & $\alpha_z\times10^4$\\
    \hline
    \hline
    R67     & $>2.4$   & $>63.58$ & $2.3^{+2.5}_{-0.9}$ & $60.93^{+66.35}_{-24.17}$ \\
    R100    & $<0.011$ & $<0.91$ & $0.0038^{+0.02}_{-0.0013}$ & $0.32^{+1.66}_{-0.12}$ \\
    \hline
    \end{tabular}
    \caption{(1) Ring ID. (2)-(3) Dust-to-gas coupling (from \citealt{Doi&Kataoka2021}) and vertical viscosity assuming the same ${\rm St}_{\rm frag}$ from \autoref{tab:1}. (4)-(5) Same as in previous columns with the dust-to-gas coupling from \citet{Liu2022}.}
    \label{tab:3}
\end{table*}

In the 67$\,$au ring, $\alpha_z\gg\alpha_r$, in \textit{qualitative} agreement with the results of \citet{Xu&Bai2022a} in the absence of significant particle loading in the mid-plane. We remark that VSI and non-ideal MHD simulations with Ohmic dead zones also predict a larger vertical than radial diffusion \citep{Schafer2020,Yang2018}. However, we determined a cooling timescale ($t_{\rm cool}\Omega_{\rm K}\approx6.5$) long enough for this ring to be stable against VSI. Additionally, we caveat that, when $\alpha_z\gg\alpha_r$, vertical turbulence might be the dominant mechanism in setting the grain size, increasing our measured $\alpha_{\rm turb}$ and $\Sigma_{\rm gas}$ by a factor of three. Nevertheless, since the estimate of a larger vertical diffusivity in R67 follows from the inference of a puffed up particle layer in this ring \citep{Doi&Kataoka2021}, we considered the possibility that this high dust scale height can also be due to dust advection in a wind, as expected from AD models (e.g., \citealt{Riols&Lesur2018}). Indeed, a molecular outflow was detected in HD~163296 and interpreted as a magneto-centrifugal wind \citep{Booth2021}. We determined the minimum wind mass loss~rate compatible with lofting 0.13$\,$mm particles to the dust scale height inferred by \citet{Doi&Kataoka2021} using the prescriptions of \citet[][Eq.~6]{Giacalone2019} and \citet[][Eq.~24]{Booth&Clarke2021}, making the conservative assumption that the wind base spans the range between 4$\,$au \citep{Booth2021} and 67$\,$au (the ring location). In the latter case we also assumed $z_{\rm IF}\approx H_{\rm IF}$ and a radially constant $\dot{\Sigma}$. Both prescriptions give similar values of $\dot{M}_{\rm loss}\approx0.5\ {\rm to}\ 4\times10^{-6}\,M_\odot\,{\rm yr}$, broadly consistent with the mass loss rate estimated by \citet{Booth2021}. However, in this scenario the thin scale height in the R100 would imply a steep decrease in the mass loss rate between 67 and 100$\,$au. In the 100$\,$au ring, instead, $\alpha_z\ll\alpha_r\approx3.32\times10^{-4}$ in \textit{quantitative} agreement with the results of \citet{Xu&Bai2022b} for particle clumping in strong pressure bumps, supporting our hypothesis that SI is underway in this ring.

To summarise, isotropic turbulence is expected to halt particle clumping under SI. However, in the rings of HD~163296 turbulence is non-isotropic. Non-ideal MHD simulations with ambipolar diffusion show that the turbulence levels in R100 are consistent with SI-driven particle clumping. However, we caveat that these models only considered particles with ${\rm St}\approx0.1$. To confirm that this process is underway in R100, new simulations with ${\rm St}\approx0.01$ particles (similar to those in \autoref{tab:1}) need to be performed.

\subsection{Particle size distribution}
All the previously mentioned simulations considered particle clumping in the monodisperse approximation. However, emission in the 67$\,$au and 100$\,$au rings of HD~163296 is consistent with a power-law particle size distribution with exponent $q=4$ \citep{Guidi2022}. Although the linear growth of SI in the polydisperse case remains debated \citep{Krapp2019}, convergence to high growth rates can be achieved when either $\epsilon\gtrsim1$ or ${\rm St}_{\rm max}\gtrsim1$, especially for more top-heavy dust size distributions \citep{Zhu&Yang2021}. Less idealised particle distributions, motivated by fragmentation-coagulation models \citep{Birnstiel2011} also seem to ease the linear growth of SI \citep{McNally2021}. Vertically stratified shearing box simulations with multiple particle species found that SI-driven particle clumping can take place \citep{Bai&Stone2010_SI,Rucska&Wadsley2023}. However, when $\max({\rm St})\lesssim10^{-2}$, especially for flatter particle size distributions, more stringent (i.e., higher $Z$) conditions than in the monodisperse case are required \citep{Schaffer2021}. Models considering how external turbulence and pressure bumps influence these results need to be performed for a proper comparison with our measurements. 

\subsection{Caveats on dust properties}
As explained in \autoref{sec:methods}, our results on the gas surface density and turbulence parameter are based on knowledge of the size, density and temperature of dust, that in turn depend on the particle composition assumed in the derivation \citep{Guidi2022}. In a recent paper, \citet{Jiang2023} estimated $\alpha_{\rm turb}$ and ${\rm St}$ radial profiles in HD~163296 extending to the entire disc our hypothesis that fragmentation limits the maximum grain size and adopting a dust-to-gas ratio of $10^{-2}$, similar to what we found in R67. Nonetheless, their results ($\alpha_{\rm turb}\approx10^{-4}$ and ${\rm St}\approx4\times10^{-2}$) are slightly different from ours because they adopted the dust density and grain size estimated by \citet{Sierra2021} using DSHARP opacities \citep{Birnstiel2018}. Assessing how our results depend on dust composition self-consistently is beyond the aim of this Letter. However, in \autoref{sec:appendix}, we provide a simple method to test how SI-driven particle clumping is affected by dust composition, under the assumption that the size, density and temperature of dust accurately recover the gas surface density estimated by \citet{Booth2019} regardless of the assumed particle opacities. We show that both DSHARP compositions and mixtures including carbonaceous material are consistent with particle clumping under SI in the 100$\,$au~ring of HD~163296. Furthermore, we discuss how our results can be used to determine a fiducial dust composition.


\subsection{Angular momentum transport}
Under the assumption that turbulence regulates angular momentum transport in the disc, we can write the kinematic disc viscosity as $\nu=\alpha_{\rm SS}c_{\rm s}^2/\Omega_{\rm K}$ \citep{Shakura&Sunyaev1973}, where $\alpha_{\rm SS}=\alpha_{\rm turb}$, and use a combination of \autoref{eq:5} and \ref{eq:6} to determine the steady-state mass accretion rate as
\begin{equation}
    \dot{M}_{\rm acc,t}=3\pi\nu\Sigma_{\rm gas}=\frac{3\pi^2}{2}\frac{c_{\rm s}^2}{\Omega_{\rm K}}\rho_{\rm s}a_{\rm frag}\left(\frac{\alpha_{\rm turb}}{{\rm St}_{\rm frag}}\right).
\end{equation}
We notice that the predicted value does not depend on the fragmentation threshold velocity. Since turbulence in the 100$\,$au ring might be affected by SI-driven particle clumping, we only determined the mass accretion rate in the 67$\,$au ring, where $\dot{M}_{\rm acc,t}=(2.21\pm0.65)\times10^{-9}\,M_\odot\,{\rm yr}^{-1}$. 

Observational measurements of the mass accretion rate of HD~163296 are debated. The most~recent estimate (obsv. date: June 2013), based on H$\alpha$ accretion luminosity, is $\log(\dot{M}_{\rm acc,obs}/M_\odot\,{\rm yr})=-6.79^{+0.15}_{-0.16}$ \citep{Wichittanakom2020}. Instead, the closest previous epoch inference (obsv. date: October 2009), from the excess emission around the Balmer jump, suggests that $\log(\dot{M}_{\rm acc,obs}/M_\odot\,{\rm yr})=-7.49^{+0.14}_{-0.30}$ \citep{Fairlamb2015}, in good agreements with same epoch estimates based on the Br$\gamma$ line luminosity \citep{Grant2023}. It is known that in HD~163296 the accretion rate experienced an abrupt increase of $\approx1\,$dex about 20 years ago (see \citealt{Mendigutia2013} and Fig.~8 of \citealt{Ellerbroek2014}). The measurement of \citet{Wichittanakom2020} is quantitatively consistent with this prolonged outburst phase (also notice that in steady-state it would imply that $\approx 0.97\,M_\odot$ of gas were accreted over the age of the system, so it is unreasonable to say that this accretion rate is maintained throughout the whole disc lifetime). The estimate of \citet{Fairlamb2015}, instead, is more consistent with mass accretion rate measured in the quiescent phase. Therefore, we would be more prone to consider the latter as fitting our steady-state picture better.

In any case, both these measurements are higher than our estimate of the turbulent mass accretion rate, implying that turbulence is not efficient enough to sustain accretion, in agreement with our hypothesis that ambipolar diffusion might be in operation in the outer disc and consistent with the detection of a magneto-centrifugal wind in the system \citep{Booth2021}. We can combine our estimate of the turbulent angular momentum transport and the observed mass accretion rate to determine the efficiency of angular momentum transport due to MHD disc winds ($\alpha_{\rm DW}$, \citealt{Suzuki2016,Tabone2022}) as
\begin{equation}
    \alpha_{\rm DW}\approx\alpha_{\rm SS}\frac{\dot{M}_{\rm acc,w}}{\dot{M}_{\rm acc,t}}=\alpha_{\rm SS}\left(\frac{\dot{M}_{\rm acc,obs}}{\dot{M}_{\rm acc,t}} - 1\right),
\end{equation}
where $\dot{M}_{\rm acc,w}$ is the mass accretion rate due to MHD winds, and the relative strength between the radial and the vertical torque ($\psi=\alpha_{\rm DW}/\alpha_{\rm SS}$, \citealt{Tabone2022}). Our results give $\alpha_{\rm DW}=4.41\times10^{-2}$, $\psi=72.31$ using $\dot{M}_{\rm acc,obs}$ of \citet{Wichittanakom2020} and $\alpha_{\rm DW}=8.30\times10^{-3}$, $\psi=13.63$ using $\dot{M}_{\rm acc,obs}$ of \citet{Fairlamb2015}.

Finally, we can relate $\alpha_{\rm DW}$ to the vertical and toroidal components of the magnetic field at the wind base, combining Eq.~3 and 6 of \citet{Tabone2022} as
\begin{equation}
    B_zB_\varphi\Bigl|_{\rm base} = \frac{3\pi}{2}\frac{\Sigma_{\rm gas}c_{\rm s}^2\alpha_{\rm DW}}{R}.
\end{equation}
Global non-ideal MHD simulations including Ohmic resistivity and ambipolar diffusion showed that $B_z\approx B_{z,0}$ is vertically constant, while $B_\varphi\approx20B_{z,0}$ at the wind base, where $B_{z,0}$ is the vertically averaged vertical component of the magnetic field \citep{Bethune2017}. In R67 $B_{z,0}=0.36\,{\rm mG}$, using $\dot{M}_{\rm acc,obs}$ of \citet{Wichittanakom2020} and $B_{z,0}=0.16\,{\rm mG}$ using $\dot{M}_{\rm acc,obs}$ of \citet{Fairlamb2015}. These estimates are consistent with the upper limits on the vertical component of the magnetic field estimated in TW~Hya \citep{Vlemmings2019} and AS~209 \citep{Harrison2021} from Zeeman splitting of the CN $J=2-1$ line hyperfine component circular polarisation observations. The corresponding plasma parameters (i.e., the ratio of the thermal to magnetic pressure) are $\beta_0=8\pi P_0B_{z,0}^{-2}=1.75\times10^4$ and $8.78\times10^4$, respectively, where $P_0\approx\Sigma_{\rm gas}c_s\Omega_{\rm K}/\sqrt{2\pi}$ is the mid-plane thermal pressure.

\section{Summary and future prospects}\label{sec:conclusion}
We introduced a new technique to measure the turbulence and gas surface density in bright disc rings, under the assumption that grain growth is limited by fragmentation at the ring centre. In the 67$\,$au and 100$\,$au ring of HD~163296, the only source where this analysis can currently be carried out, our measurements are in remarkably good agreement with independent estimates. We then combined our results with literature measurements of the dust surface density and grain size to compute the dust-to-gas surface density ratio and Stokes number. By comparison with the threshold for particle clumping under SI of \citet{Li&Youdin2021}, we found that, in the laminar case, the 100$\,$au ring is undergoing planetesimal formation. Although external isotropic turbulence could halt particle clumping, we discussed how less idealised treatments of diffusivity, consistent with evidence of non-isotropic turbulence in the rings of HD~163296, might aid SI-driven particle clumping. We proposed a consistent picture where ambipolar diffusion is operating in the outer disc, seeding SI in and favouring particle clumping in the 100$\,$au ring. This hypothesis is consistent with evidence that turbulence is not strong enough to sustain accretion in the system, in agreement with the detection of a MHD disc wind.

The recent detection of a candidate proto-planet in the 94$\,$au gap of HD~163296 \citep{Teague2018,Izquierdo2022} suggests that a first generation of planetesimals already formed in this system. Indeed, with an age of 6 Myr \citep{Wichittanakom2020}, HD~163296 is a relatively old source. In this context, our result that SI is not operating in the 67$\,$au and 100$\,$au rings, when external isotropic turbulence is considered, is not surprising and aligns with the hypothesis that planets must form early \citep{Tychoniec2020}. On the other hand, meteoritic records in the Solar System indicate that planetesimal formation could take place through the whole disc lifetime \citep{Lichtenberg2021}. If the 100$\,$au ring is unstable to SI, then it would be likely forming second generation planetesimals. Simulations resolving particle clumps and studying their accretion history are needed to assess if these planetesimals will build planetary cores (the dust mass in the 100$\,$au ring is $96^{+13}_{-16}\ M_\Earth$, see \citealt{Guidi2022}, enough to form Jupiter's core with 20\% efficiency) or a Kuiper Belt analogue.
 
Applying our newly-developed technique to a statistically significant sample spanning different disc and stellar properties (e.g., ring location, age, metallicity) could be promising to conclusively assess if SI is a viable mechanism to form planetesimals in bright rings. On the numerical side, more robust thresholds for particle clumping, exploring a parameter space consistent with our measurements (pressure maxima in equilibrium between diffusion and drift with ${\rm St}\approx0.01$ grains) are needed, but the real bottleneck is on the observational side, where our method requires high-resolution (i) emission line data to measure the dust-to-gas coupling, and (ii) multi-frequency observations to estimate the properties of dust. The exoALMA Large Program will~significantly increase the number of discs with state-of-the-art CO emission observation, ideal to study gas kinematics. However, the size, density and temperature of dust have been so far successfully constrained only in a handful of sources, mainly because emission at wavelengths longer than 3 mm, that proved crucial in this analysis (e.g., \citealt{Carrasco-Gonzalez2019}), was accessible only with VLA. We expect ALMA~Band~1 (and ngVLA in the future) to significantly expand the sample of discs with well constrained dust properties. Finally, discs where independent measurements of the gas surface density are also available can be used to further benchmark our analysis technique. Ideal sources are those targeted by the exoALMA Large Program, for gas surface density estimates based on self-gravity \citep{Veronesi2021,Lodato2023}, AGE-PRO and DECO, for gas mass estimates based on rare CO isotopologues and N$_2$H$^+$ \citep{Anderson2022,Trapman2022}.



\begin{acknowledgments}
We are grateful to the referee for their insightful comments that helped to improve the quality of our manuscript and G.~Guidi for sharing the dust density, temperature and size radial profiles of HD~163296. F.Z. acknowledges support from STFC and Cambridge Trust for a Ph.D. studentship. R.A.B. is supported by a University Research Fellowship. S.F. is funded by the European Union (ERC, UNVEIL, 101076613). G.R. is funded by the European Union under the European Union’s Horizon Europe Research \& Innovation Programme No.~101039651 (DiscEvol) and by the Fondazione Cariplo, grant no. 2022-1217. Views and opinions expressed are however those of the author(s) only and do not necessarily reflect those of the European Union or the European Research Council. Neither the European Union nor the granting authority can be held responsible for them. This project has received funding from the European Union's Horizon 2020 research and innovation programme under the Marie Sklodowska-Curie grant agreement No. 823823 (Dustbusters RISE project).
\end{acknowledgments}

%

\facilities{ALMA, VLA}


\software{\texttt{JupyterNotebook} \citep{jupyter_notebook}, \texttt{numpy} \citep{numpy}, \texttt{astropy} \citep{astropy}, \texttt{matplotlib} \citep{matplotlib}, \texttt{dsharp\_opac} \citep{Birnstiel2018}.}



\appendix

\section{The role of dust composition}\label{sec:appendix}
Our results are based on literature estimates of the temperature, density and size of dust that were obtained under the assumption of a specific solid composition (DIANA opacities, \citealt{Woitke2016,Guidi2022}). Hereafter we discuss how our results on SI-driven particle clumping are affected by this assumption. 

\begin{table*}[t!]
    \centering
    \begin{tabular}{|c|c|c|c|c|c|c|}
    \hline
    & $\max(\rho_{\rm s}a_{\rm max})$ & \multirow{2}{*}{$\max({\rm St})$} & $\max(u_{\rm frag})$ & $\min(\rho_{\rm s}a_{\rm max})$ &  \multirow{2}{*}{$\min({\rm St})$} & $\min(u_{\rm frag})$\\
    & (g$\,$cm$^{-2}$) & & (cm$\,$s$^{-1}$) & (g$\,$cm$^{-2}$) & & (cm$\,$s$^{-1}$) \\
    \hline
    \hline
    R67    & $1.59\times10^{-1}$ & $1.30\times10^{-2}$ & $4.92\times10^2$ & $3.24\times10^{-2}$ & $2.65\times10^{-3}$ & $10^2$ \\
    \hline
    R100   & $4.93\times10^{-1}$ & $7.50\times10^{-2}$ & $9.03\times10^2$ & $5.46\times10^{-2}$ & $8.31\times10^{-3}$ & $10^2$ \\
    \hline
    \end{tabular}
    \caption{Upper and lower limits on $\rho_{\rm s}a_{\rm max}$, Stokes number and fragmentation velocity threshold from \autoref{eq:A1} to~\ref{eq:A3}.}
    \label{tab:4}
\end{table*}

We can combine \autoref{eq:5} and~\ref{eq:6} to write
\begin{equation}\label{eq:A1}
    \rho_{\rm s}a_{\rm max} = \frac{2}{\pi}\Sigma_{\rm gas}\alpha_{\rm turb}\left(\frac{\alpha_{\rm turb}}{{\rm St}_{\rm frag}}\right)^{-1}.
\end{equation}
Adopting the gas surface density estimate of \citet{Booth2019}, the upper limit on turbulence of \citet{Flaherty2017} and the dust-to-gas coupling measured by \citet{Rosotti2020}, we can determine upper limits for $\rho_{\rm s}a_{\rm max}$ and the Stokes number (from \autoref{eq:4}). Our results are listed in Column (2)-(3) of \autoref{tab:4} and we stress that they do not depend on dust composition (since the right-hand side of \autoref{eq:A1} does not). 

\begin{figure*}[t!]
    \centering
    \includegraphics[width=\textwidth]{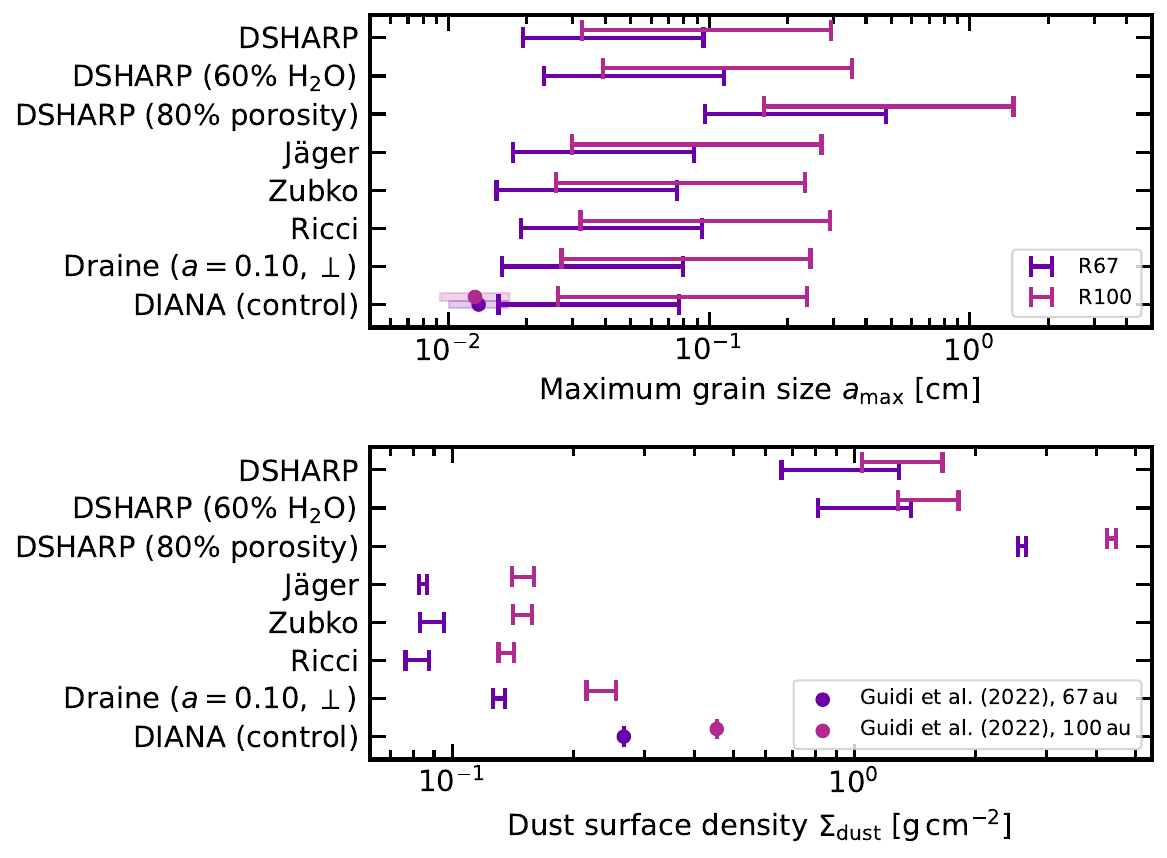}
    \caption{\textbf{Top panel:} Maximum grain sizes compatible with the gas surface density estimate of \citet{Booth2019} and the upper limit on disc turbulence of \citet{Flaherty2017}, for different dust compositions (see Appendix~B of \citealt{Birnstiel2018}) in R67 (violet) and R100 (purple). The results (best-fit and $1\sigma$ spread) of \citet{Guidi2022} are shown as dots and shaded regions of the same colour. \textbf{Bottom panel:} Dust surface density estimates based on the best-fit of \citet{Guidi2022} in the optically thin limit.}
    \label{fig:A1}
\end{figure*}

Inverting \autoref{eq:5}, we can determine the maximum dust fragmentation velocity threshold these upper limits correspond to
\begin{equation}\label{eq:A2}
    u_{\rm frag}=2.85\,\alpha_{\rm turb}c_{\rm s}\left(\frac{\alpha_{\rm turb}}{{\rm St}_{\rm frag}}\right)^{-1/2},
\end{equation}
where we estimated the isothermal sound speed adopting the best-fit temperature profile that \citet{Guidi2022} self-consistently constrained together with grain size and dust density. While this procedure might introduce some dependence of the dust temperature on the assumed composition, we consider the profile of \citet{Guidi2022} robust, since it agrees within a factor of 1.5 with independent temperature estimates based on the thermochemical models of \citet{Zhang2021} both in R67 and R100. As a consequence we expect our inferred $\max(u_{\rm frag})$ values, listed in Column (4) of \autoref{tab:4}, to be also insensitive to dust composition. We notice that $\max(u_{\rm frag})<\SI{10}{\m\per\s}$, that is often adopted as a fragmentation velocity threshold for water ice coated grains \citep{Gundlach&Blum2015}. This upper limit is consistent with the results of \citet{Gundlach2018,Musiolik&Wurm2019} and our adopted fiducial $u_{\rm frag}=\SI{1}{\m\per\s}$.

Finally, from \autoref{eq:6}
\begin{equation}\label{eq:A3}
    \rho_{\rm s}a_{\rm max} = 0.35\,\frac{2}{\pi}\Sigma_{\rm gas}\frac{u_{\rm frag}}{c_{\rm s}}\left(\frac{\alpha_{\rm turb}}{{\rm St}_{\rm frag}}\right)^{-1/2},
\end{equation}
and, under the hypothesis that $u_{\rm frag}=\SI{1}{\m\per\s}$ is the minimum fragmentation velocity threshold, we can determine lower limits for $\rho_{\rm s}a_{\rm max}$ and the Stokes number. Our results are listed in Column (5)-(6) of \autoref{tab:4} and considerations on their dependence on dust composition similar to those of \autoref{eq:A2} can be made.

We can use the upper and lower limits on $\rho_{\rm s}a_{\rm max}$ in \autoref{tab:4} to constrain the range of maximum grain sizes consistent with the gas surface density of \citet{Booth2019} and the upper limit on gas turbulence of \citet{Flaherty2017} for different dust compositions. These ranges are displayed in the top panel of \autoref{fig:A1} for R67 (violet) and R100 (purple). We took into account the same dust mixtures considered in Appendix~B of \citet{Birnstiel2018} and we refer to this paper for a detailed discussion of the materials involved. The $a_{\rm max}$ values we determined range from a few hundred microns to some millimetres, with the only exception of porous grains, whose maximum grain size can reach up to $1\,{\rm cm}$. The dots and shaded regions of the same colours display the best fit maximum grain sizes estimated by \citet{Guidi2022} and their $1\sigma$ uncertainty. Similarly to what we noticed in \autoref{sec:results}, these results are compatible within $1\sigma$ in R67 and $3\sigma$ (a factor of two) in R100, with the $\min(a_{\rm max})$ values based on the gas surface density of \citet{Booth2019} from \autoref{tab:4}. 

We stress that the grain size ranges in \autoref{fig:A1} only depend on the assumption that dust growth at the ring centre is limited by turbulent fragmentation. Therefore, they can be used to discriminate between different dust compositions by comparison with the maximum grain sizes fitted self-consistently with dust density and temperature (e.g., as \citealt{Guidi2022} did) for different dust mixtures. If the results of this analysis are not consistent with the intervals in \autoref{tab:4}, they will not be compatible with the gas surface density of \citet{Booth2019} or the turbulence upper limit of \citet{Flaherty2017}, suggesting that they are less reliable. We postpone this analysis to a future paper.

We can make a step forward hypothesising that emission is optically thin. In this case, the observed intensity can be approximated as $I_\nu\approx B_\nu(T)\kappa_\nu\Sigma_{\rm dust}$, where $B_\nu(T)$ is the Planck function at frequency $\nu$ and temperature $T$, while $\kappa_\nu$ is the dust absorption opacity at frequency $\nu$. Under the safe hypothesis that the temperature profile of \citet{Guidi2022} does not significantly change with dust composition (as discussed above), we can then compute the dust surface density for any solid composition in R67 and R100 as
\begin{equation}\label{eq:A4}
    \Sigma_{\rm dust}=\Sigma_{\rm dust,ref}\frac{\kappa_{\nu,{\rm ref}}(a_{\rm max,ref})}{\kappa_\nu(a_{\rm max})},
\end{equation}
where the subscript ``ref'' stands for the ``reference'' composition and best-fit results of \citet{Guidi2022}. We used the \texttt{dsharp\_opac} package \citep{Birnstiel2018} to generate $1.3\,{\rm mm}$ opacities for $q=4$ (as inferred by \citealt{Guidi2022} in both rings) and determined the upper and lower limits of $\Sigma_{\rm dust}$ for the minimum and maximum opacity within the range of maximum grain sizes in \autoref{tab:4} (since the opacity profile is non-monotonic, the minimum and maximum $\Sigma_{\rm dust}$ do not necessarily correspond the maximum and minimum $a_{\rm max}$). 


Our results are displayed in the bottom panel of \autoref{fig:A1}. The different optical properties of each dust mixture determine a range of dust surface densities spanning a factor of a hundred. Taking the default DSHARP opacity (``DSHARP'') as a reference, the ten times lower opacity of porous DSHARP grains (``80\% porosity'') leads to much larger densities, while increasing the water content (``60\% H$_2$O'') has only a marginal effect on our final results. Instead, very different densities are found when the organic materials typical of the DSHARP mixture are replaced by carbonaceous material, such as the ``Jäger'' \citep{Jaeger1998}, ``Zubko'' \citep{Zubko1996}, ``Ricci'' \citep{Ricci2010} and ``Draine'' \citep{Draine2003} compositions. Since they have much larger absorption opacities, they also lead to smaller dust densities. The results of \citet{Guidi2022} are intermediate between these two families.

\begin{figure*}[t!]
    \centering
    \includegraphics[width=\textwidth]{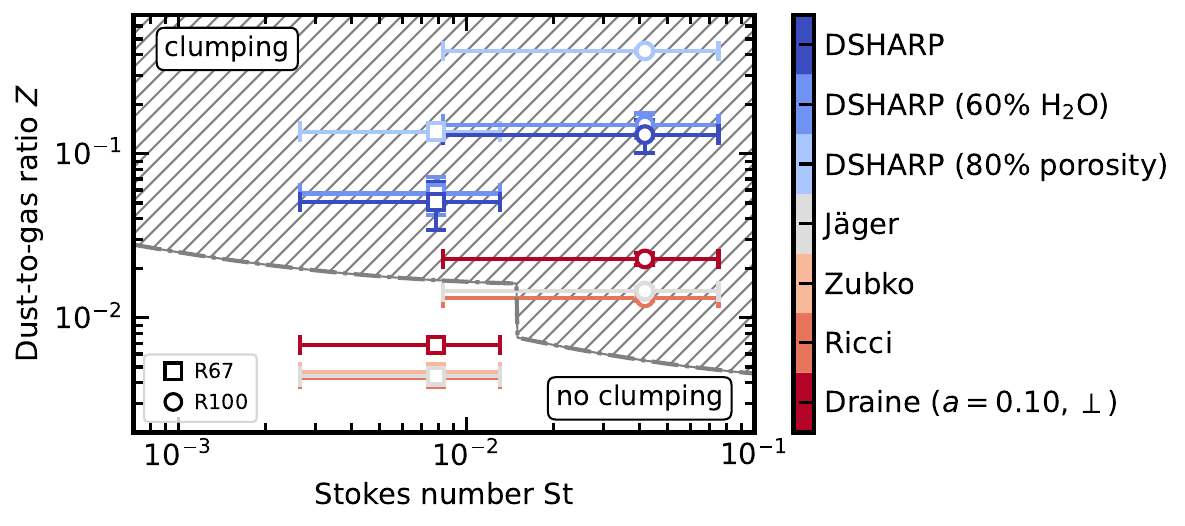}
    \caption{Comparison of the dust-to-gas surface density ratio and Stoke number for different compositions (colour-code) with the threshold for particle clumping of \citet{Li&Youdin2021} in the laminar case. Our results that the 100$\,$au ring in undergoing particle clumping under SI is consistent with different assumptions on grain mixtures.}
    \label{fig:A2}
\end{figure*}

We can use our dust surface density estimates from \autoref{eq:A4} and the gas surface density of \citet{Booth2019} to determine a range of dust-to-gas-surface density ratios for each composition. \autoref{fig:A2} shows a comparison of the threshold for particle clumping in the absence of external turbulence proposed by \citet{Li&Youdin2021} and the ($Z$, ${\rm St}$) ranges for different solid mixtures, colour-coded by dust composition. It is clear that our result that SI-driven clumping is underway in R100 is common to all the dust mixtures we tested. We also stress that it does not depend on the $q\in\{2.5,3.0,3.5,4.0\}$ and $\lambda/{\rm mm}\in\{1.3,3.1,9.1\}$ that we adopted to determine the dust densities. Instead, in R67 DSHARP opacities suggest that particle clumping is taking place, while carbonaceous grains would favour the non-clumping scenario.



\bibliography{sample631}{}

\begin{thebibliography}{}
\expandafter\ifx\csname natexlab\endcsname\relax\def\natexlab#1{#1}\fi
\providecommand{\url}[1]{\href{#1}{#1}}
\providecommand{\dodoi}[1]{doi:~\href{http://doi.org/#1}{\nolinkurl{#1}}}
\providecommand{\doeprint}[1]{\href{http://ascl.net/#1}{\nolinkurl{http://ascl.net/#1}}}
\providecommand{\doarXiv}[1]{\href{https://arxiv.org/abs/#1}{\nolinkurl{https://arxiv.org/abs/#1}}}

\bibitem[{{Anderson} {et~al.}(2022){Anderson}, {Cleeves}, {Blake}, {Bergin},
  {Zhang}, {Carpenter}, \& {Schwarz}}]{Anderson2022}
{Anderson}, D.~E., {Cleeves}, L.~I., {Blake}, G.~A., {et~al.} 2022, \apj, 927,
  229, \dodoi{10.3847/1538-4357/ac517e}

\bibitem[{{Andrews}(2020)}]{Andrews2020}
{Andrews}, S.~M. 2020, \araa, 58, 483,
  \dodoi{10.1146/annurev-astro-031220-010302}

\bibitem[{{Andrews} {et~al.}(2018){Andrews}, {Huang}, {P{\'e}rez}, {Isella},
  {Dullemond}, {Kurtovic}, {Guzm{\'a}n}, {Carpenter}, {Wilner}, {Zhang}, {Zhu},
  {Birnstiel}, {Bai}, {Benisty}, {Hughes}, {{\"O}berg}, \&
  {Ricci}}]{Andrews2018}
{Andrews}, S.~M., {Huang}, J., {P{\'e}rez}, L.~M., {et~al.} 2018, \apjl, 869,
  L41, \dodoi{10.3847/2041-8213/aaf741}

\bibitem[{{Astropy Collaboration} {et~al.}(2022){Astropy Collaboration},
  {Price-Whelan}, {Lim}, {Earl}, {Starkman}, {Bradley}, {Shupe}, {Patil},
  {Corrales}, {Brasseur}, {N{\"o}the}, {Donath}, {Tollerud}, {Morris},
  {Ginsburg}, {Vaher}, {Weaver}, {Tocknell}, {Jamieson}, {van Kerkwijk},
  {Robitaille}, {Merry}, {Bachetti}, {G{\"u}nther}, {Aldcroft},
  {Alvarado-Montes}, {Archibald}, {B{\'o}di}, {Bapat}, {Barentsen},
  {Baz{\'a}n}, {Biswas}, {Boquien}, {Burke}, {Cara}, {Cara}, {Conroy},
  {Conseil}, {Craig}, {Cross}, {Cruz}, {D'Eugenio}, {Dencheva}, {Devillepoix},
  {Dietrich}, {Eigenbrot}, {Erben}, {Ferreira}, {Foreman-Mackey}, {Fox},
  {Freij}, {Garg}, {Geda}, {Glattly}, {Gondhalekar}, {Gordon}, {Grant},
  {Greenfield}, {Groener}, {Guest}, {Gurovich}, {Handberg}, {Hart},
  {Hatfield-Dodds}, {Homeier}, {Hosseinzadeh}, {Jenness}, {Jones}, {Joseph},
  {Kalmbach}, {Karamehmetoglu}, {Ka{\l}uszy{\'n}ski}, {Kelley}, {Kern},
  {Kerzendorf}, {Koch}, {Kulumani}, {Lee}, {Ly}, {Ma}, {MacBride}, {Maljaars},
  {Muna}, {Murphy}, {Norman}, {O'Steen}, {Oman}, {Pacifici}, {Pascual},
  {Pascual-Granado}, {Patil}, {Perren}, {Pickering}, {Rastogi}, {Roulston},
  {Ryan}, {Rykoff}, {Sabater}, {Sakurikar}, {Salgado}, {Sanghi}, {Saunders},
  {Savchenko}, {Schwardt}, {Seifert-Eckert}, {Shih}, {Jain}, {Shukla}, {Sick},
  {Simpson}, {Singanamalla}, {Singer}, {Singhal}, {Sinha}, {Sip{\H{o}}cz},
  {Spitler}, {Stansby}, {Streicher}, {{\v{S}}umak}, {Swinbank}, {Taranu},
  {Tewary}, {Tremblay}, {de Val-Borro}, {Van Kooten}, {Vasovi{\'c}}, {Verma},
  {de Miranda Cardoso}, {Williams}, {Wilson}, {Winkel}, {Wood-Vasey}, {Xue},
  {Yoachim}, {Zhang}, {Zonca}, \& {Astropy Project Contributors}}]{astropy}
{Astropy Collaboration}, {Price-Whelan}, A.~M., {Lim}, P.~L., {et~al.} 2022,
  \apj, 935, 167, \dodoi{10.3847/1538-4357/ac7c74}

\bibitem[{{Bai} \& {Stone}(2010{\natexlab{a}})}]{Bai&Stone2010_pressure}
{Bai}, X.-N., \& {Stone}, J.~M. 2010{\natexlab{a}}, \apjl, 722, L220,
  \dodoi{10.1088/2041-8205/722/2/L220}

\bibitem[{{Bai} \& {Stone}(2010{\natexlab{b}})}]{Bai&Stone2010_SI}
---. 2010{\natexlab{b}}, \apj, 722, 1437, \dodoi{10.1088/0004-637X/722/2/1437}

\bibitem[{{B{\'e}thune} {et~al.}(2017){B{\'e}thune}, {Lesur}, \&
  {Ferreira}}]{Bethune2017}
{B{\'e}thune}, W., {Lesur}, G., \& {Ferreira}, J. 2017, \aap, 600, A75,
  \dodoi{10.1051/0004-6361/201630056}

\bibitem[{{Birnstiel} {et~al.}(2010){Birnstiel}, {Dullemond}, \&
  {Brauer}}]{Birnstiel2010}
{Birnstiel}, T., {Dullemond}, C.~P., \& {Brauer}, F. 2010, \aap, 513, A79,
  \dodoi{10.1051/0004-6361/200913731}

\bibitem[{{Birnstiel} {et~al.}(2016){Birnstiel}, {Fang}, \&
  {Johansen}}]{Birnstiel2016}
{Birnstiel}, T., {Fang}, M., \& {Johansen}, A. 2016, \ssr, 205, 41,
  \dodoi{10.1007/s11214-016-0256-1}

\bibitem[{{Birnstiel} {et~al.}(2012){Birnstiel}, {Klahr}, \&
  {Ercolano}}]{Birnstiel2012}
{Birnstiel}, T., {Klahr}, H., \& {Ercolano}, B. 2012, \aap, 539, A148,
  \dodoi{10.1051/0004-6361/201118136}

\bibitem[{{Birnstiel} {et~al.}(2011){Birnstiel}, {Ormel}, \&
  {Dullemond}}]{Birnstiel2011}
{Birnstiel}, T., {Ormel}, C.~W., \& {Dullemond}, C.~P. 2011, \aap, 525, A11,
  \dodoi{10.1051/0004-6361/201015228}

\bibitem[{{Birnstiel} {et~al.}(2018){Birnstiel}, {Dullemond}, {Zhu}, {Andrews},
  {Bai}, {Wilner}, {Carpenter}, {Huang}, {Isella}, {Benisty}, {P{\'e}rez}, \&
  {Zhang}}]{Birnstiel2018}
{Birnstiel}, T., {Dullemond}, C.~P., {Zhu}, Z., {et~al.} 2018, \apjl, 869, L45,
  \dodoi{10.3847/2041-8213/aaf743}

\bibitem[{{Blum}(2018)}]{Blum2018}
{Blum}, J. 2018, \ssr, 214, 52, \dodoi{10.1007/s11214-018-0486-5}

\bibitem[{{Blum} {et~al.}(2017){Blum}, {Gundlach}, {Krause}, {Fulle},
  {Johansen}, {Agarwal}, {von Borstel}, {Shi}, {Hu}, {Bentley}, {Capaccioni},
  {Colangeli}, {Della Corte}, {Fougere}, {Green}, {Ivanovski}, {Mannel},
  {Merouane}, {Migliorini}, {Rotundi}, {Schmied}, \& {Snodgrass}}]{Blum2017}
{Blum}, J., {Gundlach}, B., {Krause}, M., {et~al.} 2017, \mnras, 469, S755,
  \dodoi{10.1093/mnras/stx2741}

\bibitem[{{Booth} {et~al.}(2019){Booth}, {Walsh}, {Ilee}, {Notsu}, {Qi},
  {Nomura}, \& {Akiyama}}]{Booth2019}
{Booth}, A.~S., {Walsh}, C., {Ilee}, J.~D., {et~al.} 2019, \apjl, 882, L31,
  \dodoi{10.3847/2041-8213/ab3645}

\bibitem[{{Booth} {et~al.}(2021){Booth}, {Tabone}, {Ilee}, {Walsh}, {Aikawa},
  {Andrews}, {Bae}, {Bergin}, {Bergner}, {Bosman}, {Calahan}, {Cataldi},
  {Cleeves}, {Czekala}, {Guzm{\'a}n}, {Huang}, {Law}, {Le Gal}, {Long},
  {Loomis}, {M{\'e}nard}, {Nomura}, {{\"O}berg}, {Qi}, {Schwarz}, {Teague},
  {Tsukagoshi}, {Wilner}, {Yamato}, \& {Zhang}}]{Booth2021}
{Booth}, A.~S., {Tabone}, B., {Ilee}, J.~D., {et~al.} 2021, \apjs, 257, 16,
  \dodoi{10.3847/1538-4365/ac1ad4}

\bibitem[{{Booth} \& {Clarke}(2021)}]{Booth&Clarke2021}
{Booth}, R.~A., \& {Clarke}, C.~J. 2021, \mnras, 502, 1569,
  \dodoi{10.1093/mnras/stab090}

\bibitem[{{Brauer} {et~al.}(2008){Brauer}, {Dullemond}, \&
  {Henning}}]{Brauer2008}
{Brauer}, F., {Dullemond}, C.~P., \& {Henning}, T. 2008, \aap, 480, 859,
  \dodoi{10.1051/0004-6361:20077759}

\bibitem[{{Carrasco-Gonz{\'a}lez} {et~al.}(2019){Carrasco-Gonz{\'a}lez},
  {Sierra}, {Flock}, {Zhu}, {Henning}, {Chandler}, {Galv{\'a}n-Madrid},
  {Mac{\'\i}as}, {Anglada}, {Linz}, {Osorio}, {Rodr{\'\i}guez}, {Testi},
  {Torrelles}, {P{\'e}rez}, \& {Liu}}]{Carrasco-Gonzalez2019}
{Carrasco-Gonz{\'a}lez}, C., {Sierra}, A., {Flock}, M., {et~al.} 2019, \apj,
  883, 71, \dodoi{10.3847/1538-4357/ab3d33}

\bibitem[{{Carrera} {et~al.}(2015){Carrera}, {Johansen}, \&
  {Davies}}]{Carrera2015}
{Carrera}, D., {Johansen}, A., \& {Davies}, M.~B. 2015, \aap, 579, A43,
  \dodoi{10.1051/0004-6361/201425120}

\bibitem[{{Carrera} \& {Simon}(2022)}]{Carrera&Simon2022}
{Carrera}, D., \& {Simon}, J.~B. 2022, \apjl, 933, L10,
  \dodoi{10.3847/2041-8213/ac6b3e}

\bibitem[{{Carrera} {et~al.}(2021){Carrera}, {Simon}, {Li}, {Kretke}, \&
  {Klahr}}]{Carrera2021}
{Carrera}, D., {Simon}, J.~B., {Li}, R., {Kretke}, K.~A., \& {Klahr}, H. 2021,
  \aj, 161, 96, \dodoi{10.3847/1538-3881/abd4d9}

\bibitem[{{Carrera} {et~al.}(2022){Carrera}, {Thomas}, {Simon}, {Small},
  {Kretke}, \& {Klahr}}]{Carrera2022}
{Carrera}, D., {Thomas}, A.~J., {Simon}, J.~B., {et~al.} 2022, \apj, 927, 52,
  \dodoi{10.3847/1538-4357/ac4d28}

\bibitem[{{Doi} \& {Kataoka}(2021)}]{Doi&Kataoka2021}
{Doi}, K., \& {Kataoka}, A. 2021, \apj, 912, 164,
  \dodoi{10.3847/1538-4357/abe5a6}

\bibitem[{{Doi} \& {Kataoka}(2023)}]{Doi&Kataoka2023}
---. 2023, arXiv e-prints, arXiv:2308.16574, \dodoi{10.48550/arXiv.2308.16574}

\bibitem[{{Dorschner} {et~al.}(1995){Dorschner}, {Begemann}, {Henning},
  {Jaeger}, \& {Mutschke}}]{Dorschner1995}
{Dorschner}, J., {Begemann}, B., {Henning}, T., {Jaeger}, C., \& {Mutschke}, H.
  1995, \aap, 300, 503

\bibitem[{{Draine}(2003)}]{Draine2003}
{Draine}, B.~T. 2003, \araa, 41, 241,
  \dodoi{10.1146/annurev.astro.41.011802.094840}

\bibitem[{{Dr{\k{a}}{\.z}kowska} {et~al.}(2023){Dr{\k{a}}{\.z}kowska},
  {Bitsch}, {Lambrechts}, {Mulders}, {Harsono}, {Vazan}, {Liu}, {Ormel},
  {Kretke}, \& {Morbidelli}}]{Drazkowska_ppvii}
{Dr{\k{a}}{\.z}kowska}, J., {Bitsch}, B., {Lambrechts}, M., {et~al.} 2023, in
  Astronomical Society of the Pacific Conference Series, Vol. 534, Astronomical
  Society of the Pacific Conference Series, ed. S.~{Inutsuka}, Y.~{Aikawa},
  T.~{Muto}, K.~{Tomida}, \& M.~{Tamura}, 717,
  \dodoi{10.48550/arXiv.2203.09759}

\bibitem[{{Dullemond} {et~al.}(2018){Dullemond}, {Birnstiel}, {Huang},
  {Kurtovic}, {Andrews}, {Guzm{\'a}n}, {P{\'e}rez}, {Isella}, {Zhu}, {Benisty},
  {Wilner}, {Bai}, {Carpenter}, {Zhang}, \& {Ricci}}]{Dullemond2018}
{Dullemond}, C.~P., {Birnstiel}, T., {Huang}, J., {et~al.} 2018, \apjl, 869,
  L46, \dodoi{10.3847/2041-8213/aaf742}

\bibitem[{{Ellerbroek} {et~al.}(2014){Ellerbroek}, {Podio}, {Dougados},
  {Cabrit}, {Sitko}, {Sana}, {Kaper}, {de Koter}, {Klaassen}, {Mulders},
  {Mendigut{\'\i}a}, {Grady}, {Grankin}, {van Winckel}, {Bacciotti}, {Russell},
  {Lynch}, {Hammel}, {Beerman}, {Day}, {Huelsman}, {Werren}, {Henden}, \&
  {Grindlay}}]{Ellerbroek2014}
{Ellerbroek}, L.~E., {Podio}, L., {Dougados}, C., {et~al.} 2014, \aap, 563,
  A87, \dodoi{10.1051/0004-6361/201323092}

\bibitem[{{Fairlamb} {et~al.}(2015){Fairlamb}, {Oudmaijer}, {Mendigut{\'\i}a},
  {Ilee}, \& {van den Ancker}}]{Fairlamb2015}
{Fairlamb}, J.~R., {Oudmaijer}, R.~D., {Mendigut{\'\i}a}, I., {Ilee}, J.~D., \&
  {van den Ancker}, M.~E. 2015, \mnras, 453, 976, \dodoi{10.1093/mnras/stv1576}

\bibitem[{{Flaherty} {et~al.}(2017){Flaherty}, {Hughes}, {Rose}, {Simon}, {Qi},
  {Andrews}, {K{\'o}sp{\'a}l}, {Wilner}, {Chiang}, {Armitage}, \&
  {Bai}}]{Flaherty2017}
{Flaherty}, K.~M., {Hughes}, A.~M., {Rose}, S.~C., {et~al.} 2017, \apj, 843,
  150, \dodoi{10.3847/1538-4357/aa79f9}

\bibitem[{{Gaia Collaboration} {et~al.}(2021){Gaia Collaboration}, {Brown},
  {Vallenari}, {Prusti}, {de Bruijne}, {Babusiaux}, {Biermann}, {Creevey},
  {Evans}, {Eyer}, {Hutton}, {Jansen}, {Jordi}, {Klioner}, {Lammers},
  {Lindegren}, {Luri}, {Mignard}, {Panem}, {Pourbaix}, {Randich}, {Sartoretti},
  {Soubiran}, {Walton}, {Arenou}, {Bailer-Jones}, {Bastian}, {Cropper},
  {Drimmel}, {Katz}, {Lattanzi}, {van Leeuwen}, {Bakker}, {Cacciari},
  {Casta{\~n}eda}, {De Angeli}, {Ducourant}, {Fabricius}, {Fouesneau},
  {Fr{\'e}mat}, {Guerra}, {Guerrier}, {Guiraud}, {Jean-Antoine Piccolo},
  {Masana}, {Messineo}, {Mowlavi}, {Nicolas}, {Nienartowicz}, {Pailler},
  {Panuzzo}, {Riclet}, {Roux}, {Seabroke}, {Sordo}, {Tanga}, {Th{\'e}venin},
  {Gracia-Abril}, {Portell}, {Teyssier}, {Altmann}, {Andrae}, {Bellas-Velidis},
  {Benson}, {Berthier}, {Blomme}, {Brugaletta}, {Burgess}, {Busso}, {Carry},
  {Cellino}, {Cheek}, {Clementini}, {Damerdji}, {Davidson}, {Delchambre},
  {Dell'Oro}, {Fern{\'a}ndez-Hern{\'a}ndez}, {Galluccio}, {Garc{\'\i}a-Lario},
  {Garcia-Reinaldos}, {Gonz{\'a}lez-N{\'u}{\~n}ez}, {Gosset}, {Haigron},
  {Halbwachs}, {Hambly}, {Harrison}, {Hatzidimitriou}, {Heiter},
  {Hern{\'a}ndez}, {Hestroffer}, {Hodgkin}, {Holl}, {Jan{\ss}en}, {Jevardat de
  Fombelle}, {Jordan}, {Krone-Martins}, {Lanzafame}, {L{\"o}ffler}, {Lorca},
  {Manteiga}, {Marchal}, {Marrese}, {Moitinho}, {Mora}, {Muinonen}, {Osborne},
  {Pancino}, {Pauwels}, {Petit}, {Recio-Blanco}, {Richards}, {Riello},
  {Rimoldini}, {Robin}, {Roegiers}, {Rybizki}, {Sarro}, {Siopis}, {Smith},
  {Sozzetti}, {Ulla}, {Utrilla}, {van Leeuwen}, {van Reeven}, {Abbas}, {Abreu
  Aramburu}, {Accart}, {Aerts}, {Aguado}, {Ajaj}, {Altavilla}, {{\'A}lvarez},
  {{\'A}lvarez Cid-Fuentes}, {Alves}, {Anderson}, {Anglada Varela}, {Antoja},
  {Audard}, {Baines}, {Baker}, {Balaguer-N{\'u}{\~n}ez}, {Balbinot}, {Balog},
  {Barache}, {Barbato}, {Barros}, {Barstow}, {Bartolom{\'e}}, {Bassilana},
  {Bauchet}, {Baudesson-Stella}, {Becciani}, {Bellazzini}, {Bernet}, {Bertone},
  {Bianchi}, {Blanco-Cuaresma}, {Boch}, {Bombrun}, {Bossini}, {Bouquillon},
  {Bragaglia}, {Bramante}, {Breedt}, {Bressan}, {Brouillet}, {Bucciarelli},
  {Burlacu}, {Busonero}, {Butkevich}, {Buzzi}, {Caffau}, {Cancelliere},
  {C{\'a}novas}, {Cantat-Gaudin}, {Carballo}, {Carlucci}, {Carnerero},
  {Carrasco}, {Casamiquela}, {Castellani}, {Castro-Ginard}, {Castro Sampol},
  {Chaoul}, {Charlot}, {Chemin}, {Chiavassa}, {Cioni}, {Comoretto}, {Cooper},
  {Cornez}, {Cowell}, {Crifo}, {Crosta}, {Crowley}, {Dafonte}, {Dapergolas},
  {David}, {David}, {de Laverny}, {De Luise}, {De March}, {De Ridder}, {de
  Souza}, {de Teodoro}, {de Torres}, {del Peloso}, {del Pozo}, {Delbo},
  {Delgado}, {Delgado}, {Delisle}, {Di Matteo}, {Diakite}, {Diener},
  {Distefano}, {Dolding}, {Eappachen}, {Edvardsson}, {Enke}, {Esquej}, {Fabre},
  {Fabrizio}, {Faigler}, {Fedorets}, {Fernique}, {Fienga}, {Figueras},
  {Fouron}, {Fragkoudi}, {Fraile}, {Franke}, {Gai}, {Garabato},
  {Garcia-Gutierrez}, {Garc{\'\i}a-Torres}, {Garofalo}, {Gavras}, {Gerlach},
  {Geyer}, {Giacobbe}, {Gilmore}, {Girona}, {Giuffrida}, {Gomel}, {Gomez},
  {Gonzalez-Santamaria}, {Gonz{\'a}lez-Vidal}, {Granvik},
  {Guti{\'e}rrez-S{\'a}nchez}, {Guy}, {Hauser}, {Haywood}, {Helmi}, {Hidalgo},
  {Hilger}, {H{\l}adczuk}, {Hobbs}, {Holland}, {Huckle}, {Jasniewicz},
  {Jonker}, {Juaristi Campillo}, {Julbe}, {Karbevska}, {Kervella}, {Khanna},
  {Kochoska}, {Kontizas}, {Kordopatis}, {Korn}, {Kostrzewa-Rutkowska},
  {Kruszy{\'n}ska}, {Lambert}, {Lanza}, {Lasne}, {Le Campion}, {Le Fustec},
  {Lebreton}, {Lebzelter}, {Leccia}, {Leclerc}, {Lecoeur-Taibi}, {Liao},
  {Licata}, {Lindstr{\o}m}, {Lister}, {Livanou}, {Lobel}, {Madrero Pardo},
  {Managau}, {Mann}, {Marchant}, {Marconi}, {Marcos Santos}, {Marinoni},
  {Marocco}, {Marshall}, {Martin Polo}, {Mart{\'\i}n-Fleitas}, {Masip},
  {Massari}, {Mastrobuono-Battisti}, {Mazeh}, {McMillan}, {Messina},
  {Michalik}, {Millar}, {Mints}, {Molina}, {Molinaro}, {Moln{\'a}r},
  {Montegriffo}, {Mor}, {Morbidelli}, {Morel}, {Morris}, {Mulone}, {Munoz},
  {Muraveva}, {Murphy}, {Musella}, {Noval}, {Ord{\'e}novic}, {Orr{\`u}},
  {Osinde}, {Pagani}, {Pagano}, {Palaversa}, {Palicio}, {Panahi}, {Pawlak},
  {Pe{\~n}alosa Esteller}, {Penttil{\"a}}, {Piersimoni}, {Pineau}, {Plachy},
  {Plum}, {Poggio}, {Poretti}, {Poujoulet}, {Pr{\v{s}}a}, {Pulone}, {Racero},
  {Ragaini}, {Rainer}, {Raiteri}, {Rambaux}, {Ramos}, {Ramos-Lerate}, {Re
  Fiorentin}, {Regibo}, {Reyl{\'e}}, {Ripepi}, {Riva}, {Rixon}, {Robichon},
  {Robin}, {Roelens}, {Rohrbasser}, {Romero-G{\'o}mez}, {Rowell}, {Royer},
  {Rybicki}, {Sadowski}, {Sagrist{\`a} Sell{\'e}s}, {Sahlmann}, {Salgado},
  {Salguero}, {Samaras}, {Sanchez Gimenez}, {Sanna}, {Santove{\~n}a},
  {Sarasso}, {Schultheis}, {Sciacca}, {Segol}, {Segovia}, {S{\'e}gransan},
  {Semeux}, {Shahaf}, {Siddiqui}, {Siebert}, {Siltala}, {Slezak}, {Smart},
  {Solano}, {Solitro}, {Souami}, {Souchay}, {Spagna}, {Spoto}, {Steele},
  {Steidelm{\"u}ller}, {Stephenson}, {S{\"u}veges}, {Szabados}, {Szegedi-Elek},
  {Taris}, {Tauran}, {Taylor}, {Teixeira}, {Thuillot}, {Tonello}, {Torra},
  {Torra}, {Turon}, {Unger}, {Vaillant}, {van Dillen}, {Vanel}, {Vecchiato},
  {Viala}, {Vicente}, {Voutsinas}, {Weiler}, {Wevers}, {Wyrzykowski}, {Yoldas},
  {Yvard}, {Zhao}, {Zorec}, {Zucker}, {Zurbach}, \& {Zwitter}}]{Gaia_EDR3}
{Gaia Collaboration}, {Brown}, A.~G.~A., {Vallenari}, A., {et~al.} 2021, \aap,
  649, A1, \dodoi{10.1051/0004-6361/202039657}

\bibitem[{{Giacalone} {et~al.}(2019){Giacalone}, {Teitler}, {K{\"o}nigl},
  {Krijt}, \& {Ciesla}}]{Giacalone2019}
{Giacalone}, S., {Teitler}, S., {K{\"o}nigl}, A., {Krijt}, S., \& {Ciesla},
  F.~J. 2019, \apj, 882, 33, \dodoi{10.3847/1538-4357/ab311a}

\bibitem[{{Gole} {et~al.}(2020){Gole}, {Simon}, {Li}, {Youdin}, \&
  {Armitage}}]{Gole2020}
{Gole}, D.~A., {Simon}, J.~B., {Li}, R., {Youdin}, A.~N., \& {Armitage}, P.~J.
  2020, \apj, 904, 132, \dodoi{10.3847/1538-4357/abc334}

\bibitem[{{Grant} {et~al.}(2023){Grant}, {Stapper}, {Hogerheijde}, {van
  Dishoeck}, {Brittain}, \& {Vioque}}]{Grant2023}
{Grant}, S.~L., {Stapper}, L.~M., {Hogerheijde}, M.~R., {et~al.} 2023, \aj,
  166, 147, \dodoi{10.3847/1538-3881/acf128}

\bibitem[{{Guidi} {et~al.}(2022){Guidi}, {Isella}, {Testi}, {Chandler}, {Liu},
  {Schmid}, {Rosotti}, {Meng}, {Jennings}, {Williams}, {Carpenter}, {de
  Gregorio-Monsalvo}, {Li}, {Liu}, {Ortolani}, {Quanz}, {Ricci}, \&
  {Tazzari}}]{Guidi2022}
{Guidi}, G., {Isella}, A., {Testi}, L., {et~al.} 2022, \aap, 664, A137,
  \dodoi{10.1051/0004-6361/202142303}

\bibitem[{{Gundlach} \& {Blum}(2015)}]{Gundlach&Blum2015}
{Gundlach}, B., \& {Blum}, J. 2015, \apj, 798, 34,
  \dodoi{10.1088/0004-637X/798/1/34}

\bibitem[{{Gundlach} {et~al.}(2018){Gundlach}, {Schmidt}, {Kreuzig},
  {Bischoff}, {Rezaei}, {Kothe}, {Blum}, {Grzesik}, \& {Stoll}}]{Gundlach2018}
{Gundlach}, B., {Schmidt}, K.~P., {Kreuzig}, C., {et~al.} 2018, \mnras, 479,
  1273, \dodoi{10.1093/mnras/sty1550}

\bibitem[{{Harris} {et~al.}(2020){Harris}, {Millman}, {van der Walt},
  {Gommers}, {Virtanen}, {Cournapeau}, {Wieser}, {Taylor}, {Berg}, {Smith},
  {Kern}, {Picus}, {Hoyer}, {van Kerkwijk}, {Brett}, {Haldane}, {del R{\'\i}o},
  {Wiebe}, {Peterson}, {G{\'e}rard-Marchant}, {Sheppard}, {Reddy}, {Weckesser},
  {Abbasi}, {Gohlke}, \& {Oliphant}}]{numpy}
{Harris}, C.~R., {Millman}, K.~J., {van der Walt}, S.~J., {et~al.} 2020, \nat,
  585, 357, \dodoi{10.1038/s41586-020-2649-2}

\bibitem[{{Harrison} {et~al.}(2021){Harrison}, {Looney}, {Stephens}, {Li},
  {Teague}, {Crutcher}, {Yang}, {Cox}, {Fern{\'a}ndez-L{\'o}pez}, \&
  {Shinnaga}}]{Harrison2021}
{Harrison}, R.~E., {Looney}, L.~W., {Stephens}, I.~W., {et~al.} 2021, \apj,
  908, 141, \dodoi{10.3847/1538-4357/abd94e}

\bibitem[{Hunter(2007)}]{matplotlib}
Hunter, J.~D. 2007, Computing in Science \& Engineering, 9, 90,
  \dodoi{10.1109/MCSE.2007.55}

\bibitem[{{Isella} {et~al.}(2018){Isella}, {Huang}, {Andrews}, {Dullemond},
  {Birnstiel}, {Zhang}, {Zhu}, {Guzm{\'a}n}, {P{\'e}rez}, {Bai}, {Benisty},
  {Carpenter}, {Ricci}, \& {Wilner}}]{Isella2018}
{Isella}, A., {Huang}, J., {Andrews}, S.~M., {et~al.} 2018, \apjl, 869, L49,
  \dodoi{10.3847/2041-8213/aaf747}

\bibitem[{{Izquierdo} {et~al.}(2022){Izquierdo}, {Facchini}, {Rosotti}, {van
  Dishoeck}, \& {Testi}}]{Izquierdo2022}
{Izquierdo}, A.~F., {Facchini}, S., {Rosotti}, G.~P., {van Dishoeck}, E.~F., \&
  {Testi}, L. 2022, \apj, 928, 2, \dodoi{10.3847/1538-4357/ac474d}

\bibitem[{{Izquierdo} {et~al.}(2023){Izquierdo}, {Testi}, {Facchini},
  {Rosotti}, {van Dishoeck}, {W{\"o}lfer}, \&
  {Paneque-Carre{\~n}o}}]{Izquierdo2023}
{Izquierdo}, A.~F., {Testi}, L., {Facchini}, S., {et~al.} 2023, \aap, 674,
  A113, \dodoi{10.1051/0004-6361/202245425}

\bibitem[{{J{\"a}ger} {et~al.}(1998){J{\"a}ger}, {Mutschke}, \&
  {Henning}}]{Jaeger1998}
{J{\"a}ger}, C., {Mutschke}, H., \& {Henning}, T. 1998, \aap, 332, 291

\bibitem[{{Jiang} {et~al.}(2023){Jiang}, {Mac{\'\i}as}, {Guerra-Alvarado}, \&
  {Carrasco-Gonz{\'a}lez}}]{Jiang2023}
{Jiang}, H., {Mac{\'\i}as}, E., {Guerra-Alvarado}, O.~M., \&
  {Carrasco-Gonz{\'a}lez}, C. 2023, arXiv e-prints, arXiv:2311.07775.
\newblock \doarXiv{2311.07775}

\bibitem[{{Johansen} {et~al.}(2014){Johansen}, {Blum}, {Tanaka}, {Ormel},
  {Bizzarro}, \& {Rickman}}]{Johansen_ppvi}
{Johansen}, A., {Blum}, J., {Tanaka}, H., {et~al.} 2014, in Protostars and
  Planets VI, ed. H.~{Beuther}, R.~S. {Klessen}, C.~P. {Dullemond}, \&
  T.~{Henning}, 547--570, \dodoi{10.2458/azu_uapress_9780816531240-ch024}

\bibitem[{{Johansen} {et~al.}(2007){Johansen}, {Oishi}, {Mac Low}, {Klahr},
  {Henning}, \& {Youdin}}]{Johansen2007}
{Johansen}, A., {Oishi}, J.~S., {Mac Low}, M.-M., {et~al.} 2007, \nat, 448,
  1022, \dodoi{10.1038/nature06086}

\bibitem[{{Johansen} \& {Youdin}(2007)}]{Johansen&Youdin2007}
{Johansen}, A., \& {Youdin}, A. 2007, \apj, 662, 627, \dodoi{10.1086/516730}

\bibitem[{{Johansen} {et~al.}(2009){Johansen}, {Youdin}, \& {Mac
  Low}}]{Johansen2009}
{Johansen}, A., {Youdin}, A., \& {Mac Low}, M.-M. 2009, \apjl, 704, L75,
  \dodoi{10.1088/0004-637X/704/2/L75}

\bibitem[{{Kavelaars} {et~al.}(2021){Kavelaars}, {Petit}, {Gladman},
  {Bannister}, {Alexandersen}, {Chen}, {Gwyn}, \& {Volk}}]{Kavelaars2021}
{Kavelaars}, J.~J., {Petit}, J.-M., {Gladman}, B., {et~al.} 2021, \apjl, 920,
  L28, \dodoi{10.3847/2041-8213/ac2c72}

\bibitem[{Kluyver {et~al.}(2016)Kluyver, Ragan-Kelley, P{\'e}rez, Granger,
  Bussonnier, Frederic, Kelley, Hamrick, Grout, Corlay, Ivanov, Avila, Abdalla,
  Willing, \& development team}]{jupyter_notebook}
Kluyver, T., Ragan-Kelley, B., P{\'e}rez, F., {et~al.} 2016, in Positioning and
  Power in Academic Publishing: Players, Agents and Agendas, ed. F.~Loizides \&
  B.~Scmidt (IOS Press), 87--90.
\newblock \url{https://eprints.soton.ac.uk/403913/}

\bibitem[{{Krapp} {et~al.}(2019){Krapp}, {Ben{\'\i}tez-Llambay}, {Gressel}, \&
  {Pessah}}]{Krapp2019}
{Krapp}, L., {Ben{\'\i}tez-Llambay}, P., {Gressel}, O., \& {Pessah}, M.~E.
  2019, \apjl, 878, L30, \dodoi{10.3847/2041-8213/ab2596}

\bibitem[{{Lesur} {et~al.}(2023){Lesur}, {Flock}, {Ercolano}, {Lin}, {Yang},
  {Barranco}, {Benitez-Llambay}, {Goodman}, {Johansen}, {Klahr}, {Laibe},
  {Lyra}, {Marcus}, {Nelson}, {Squire}, {Simon}, {Turner}, {Umurhan}, \&
  {Youdin}}]{Lesur_ppvii}
{Lesur}, G., {Flock}, M., {Ercolano}, B., {et~al.} 2023, in Astronomical
  Society of the Pacific Conference Series, Vol. 534, Astronomical Society of
  the Pacific Conference Series, ed. S.~{Inutsuka}, Y.~{Aikawa}, T.~{Muto},
  K.~{Tomida}, \& M.~{Tamura}, 465

\bibitem[{{Li} \& {Youdin}(2021)}]{Li&Youdin2021}
{Li}, R., \& {Youdin}, A.~N. 2021, \apj, 919, 107,
  \dodoi{10.3847/1538-4357/ac0e9f}

\bibitem[{{Lichtenberg} {et~al.}(2021){Lichtenberg}, {Dr{\k{a}}{\.z}kowska},
  {Sch{\"o}nb{\"a}chler}, {Golabek}, \& {Hands}}]{Lichtenberg2021}
{Lichtenberg}, T., {Dr{\k{a}}{\.z}kowska}, J., {Sch{\"o}nb{\"a}chler}, M.,
  {Golabek}, G.~J., \& {Hands}, T.~O. 2021, Science, 371, 365,
  \dodoi{10.1126/science.abb3091}

\bibitem[{{Liu} {et~al.}(2022){Liu}, {Bertrang}, {Flock}, {Rosotti}, {van
  Dishoeck}, {Boehler}, {Facchini}, {Cui}, {Wolf}, \& {Fang}}]{Liu2022}
{Liu}, Y., {Bertrang}, G. H.~M., {Flock}, M., {et~al.} 2022, Science China
  Physics, Mechanics, and Astronomy, 65, 129511,
  \dodoi{10.1007/s11433-022-1982-y}

\bibitem[{{Lodato} {et~al.}(2023){Lodato}, {Rampinelli}, {Viscardi},
  {Longarini}, {Izquierdo}, {Paneque-Carre{\~n}o}, {Testi}, {Facchini},
  {Miotello}, {Veronesi}, \& {Hall}}]{Lodato2023}
{Lodato}, G., {Rampinelli}, L., {Viscardi}, E., {et~al.} 2023, \mnras, 518,
  4481, \dodoi{10.1093/mnras/stac3223}

\bibitem[{{Long} {et~al.}(2018){Long}, {Pinilla}, {Herczeg}, {Harsono},
  {Dipierro}, {Pascucci}, {Hendler}, {Tazzari}, {Ragusa}, {Salyk}, {Edwards},
  {Lodato}, {van de Plas}, {Johnstone}, {Liu}, {Boehler}, {Cabrit}, {Manara},
  {Menard}, {Mulders}, {Nisini}, {Fischer}, {Rigliaco}, {Banzatti}, {Avenhaus},
  \& {Gully-Santiago}}]{Long2018}
{Long}, F., {Pinilla}, P., {Herczeg}, G.~J., {et~al.} 2018, \apj, 869, 17,
  \dodoi{10.3847/1538-4357/aae8e1}

\bibitem[{{Mac{\'\i}as} {et~al.}(2021){Mac{\'\i}as}, {Guerra-Alvarado},
  {Carrasco-Gonz{\'a}lez}, {Ribas}, {Espaillat}, {Huang}, \&
  {Andrews}}]{Macias2021}
{Mac{\'\i}as}, E., {Guerra-Alvarado}, O., {Carrasco-Gonz{\'a}lez}, C., {et~al.}
  2021, \aap, 648, A33, \dodoi{10.1051/0004-6361/202039812}

\bibitem[{{McKinnon} {et~al.}(2020){McKinnon}, {Richardson}, {Marohnic},
  {Keane}, {Grundy}, {Hamilton}, {Nesvorn{\'y}}, {Umurhan}, {Lauer}, {Singer},
  {Stern}, {Weaver}, {Spencer}, {Buie}, {Moore}, {Kavelaars}, {Lisse}, {Mao},
  {Parker}, {Porter}, {Showalter}, {Olkin}, {Cruikshank}, {Elliott},
  {Gladstone}, {Parker}, {Verbiscer}, {Young}, \& {New Horizons Science
  Team}}]{McKinnon2020}
{McKinnon}, W.~B., {Richardson}, D.~C., {Marohnic}, J.~C., {et~al.} 2020,
  Science, 367, aay6620, \dodoi{10.1126/science.aay6620}

\bibitem[{{McNally} {et~al.}(2021){McNally}, {Lovascio}, \&
  {Paardekooper}}]{McNally2021}
{McNally}, C.~P., {Lovascio}, F., \& {Paardekooper}, S.-J. 2021, \mnras, 502,
  1469, \dodoi{10.1093/mnras/stab112}

\bibitem[{{Mendigut{\'\i}a} {et~al.}(2013){Mendigut{\'\i}a}, {Brittain},
  {Eiroa}, {Meeus}, {Montesinos}, {Mora}, {Muzerolle}, {Oudmaijer}, \&
  {Rigliaco}}]{Mendigutia2013}
{Mendigut{\'\i}a}, I., {Brittain}, S., {Eiroa}, C., {et~al.} 2013, \apj, 776,
  44, \dodoi{10.1088/0004-637X/776/1/44}

\bibitem[{{Musiolik} \& {Wurm}(2019)}]{Musiolik&Wurm2019}
{Musiolik}, G., \& {Wurm}, G. 2019, \apj, 873, 58,
  \dodoi{10.3847/1538-4357/ab0428}

\bibitem[{{Nakagawa} {et~al.}(1986){Nakagawa}, {Sekiya}, \&
  {Hayashi}}]{Nakagawa1986}
{Nakagawa}, Y., {Sekiya}, M., \& {Hayashi}, C. 1986, \icarus, 67, 375,
  \dodoi{10.1016/0019-1035(86)90121-1}

\bibitem[{{Nesvorn{\'y}} {et~al.}(2019){Nesvorn{\'y}}, {Li}, {Youdin}, {Simon},
  \& {Grundy}}]{Nesvorny2019}
{Nesvorn{\'y}}, D., {Li}, R., {Youdin}, A.~N., {Simon}, J.~B., \& {Grundy},
  W.~M. 2019, Nature Astronomy, 3, 808, \dodoi{10.1038/s41550-019-0806-z}

\bibitem[{{{\"O}berg} {et~al.}(2021){{\"O}berg}, {Guzm{\'a}n}, {Walsh},
  {Aikawa}, {Bergin}, {Law}, {Loomis}, {Alarc{\'o}n}, {Andrews}, {Bae},
  {Bergner}, {Boehler}, {Booth}, {Bosman}, {Calahan}, {Cataldi}, {Cleeves},
  {Czekala}, {Furuya}, {Huang}, {Ilee}, {Kurtovic}, {Le Gal}, {Liu}, {Long},
  {M{\'e}nard}, {Nomura}, {P{\'e}rez}, {Qi}, {Schwarz}, {Sierra}, {Teague},
  {Tsukagoshi}, {Yamato}, {van't Hoff}, {Waggoner}, {Wilner}, \&
  {Zhang}}]{Oberg2021}
{{\"O}berg}, K.~I., {Guzm{\'a}n}, V.~V., {Walsh}, C., {et~al.} 2021, \apjs,
  257, 1, \dodoi{10.3847/1538-4365/ac1432}

\bibitem[{{Ormel} \& {Cuzzi}(2007)}]{Ormel&Cuzzi2007}
{Ormel}, C.~W., \& {Cuzzi}, J.~N. 2007, \aap, 466, 413,
  \dodoi{10.1051/0004-6361:20066899}

\bibitem[{{Ricci} {et~al.}(2010){Ricci}, {Testi}, {Natta}, {Neri}, {Cabrit}, \&
  {Herczeg}}]{Ricci2010}
{Ricci}, L., {Testi}, L., {Natta}, A., {et~al.} 2010, \aap, 512, A15,
  \dodoi{10.1051/0004-6361/200913403}

\bibitem[{{Riols} \& {Lesur}(2018)}]{Riols&Lesur2018}
{Riols}, A., \& {Lesur}, G. 2018, \aap, 617, A117,
  \dodoi{10.1051/0004-6361/201833212}

\bibitem[{{Rosotti} {et~al.}(2020){Rosotti}, {Teague}, {Dullemond}, {Booth}, \&
  {Clarke}}]{Rosotti2020}
{Rosotti}, G.~P., {Teague}, R., {Dullemond}, C., {Booth}, R.~A., \& {Clarke},
  C.~J. 2020, \mnras, 495, 173, \dodoi{10.1093/mnras/staa1170}

\bibitem[{{Rucska} \& {Wadsley}(2023)}]{Rucska&Wadsley2023}
{Rucska}, J.~J., \& {Wadsley}, J.~W. 2023, \mnras, 526, 1757,
  \dodoi{10.1093/mnras/stad2855}

\bibitem[{{Scardoni} {et~al.}(2021){Scardoni}, {Booth}, \&
  {Clarke}}]{Scardoni2021}
{Scardoni}, C.~E., {Booth}, R.~A., \& {Clarke}, C.~J. 2021, \mnras, 504, 1495,
  \dodoi{10.1093/mnras/stab854}

\bibitem[{{Sch{\"a}fer} {et~al.}(2020){Sch{\"a}fer}, {Johansen}, \&
  {Banerjee}}]{Schafer2020}
{Sch{\"a}fer}, U., {Johansen}, A., \& {Banerjee}, R. 2020, \aap, 635, A190,
  \dodoi{10.1051/0004-6361/201937371}

\bibitem[{{Schaffer} {et~al.}(2021){Schaffer}, {Johansen}, \&
  {Lambrechts}}]{Schaffer2021}
{Schaffer}, N., {Johansen}, A., \& {Lambrechts}, M. 2021, \aap, 653, A14,
  \dodoi{10.1051/0004-6361/202140690}

\bibitem[{{Sekiya} \& {Onishi}(2018)}]{Sekiya&Onishi2018}
{Sekiya}, M., \& {Onishi}, I.~K. 2018, \apj, 860, 140,
  \dodoi{10.3847/1538-4357/aac4a7}

\bibitem[{{Shakura} \& {Sunyaev}(1973)}]{Shakura&Sunyaev1973}
{Shakura}, N.~I., \& {Sunyaev}, R.~A. 1973, \aap, 24, 337

\bibitem[{{Sierra} {et~al.}(2021){Sierra}, {P{\'e}rez}, {Zhang}, {Law},
  {Guzm{\'a}n}, {Qi}, {Bosman}, {{\"O}berg}, {Andrews}, {Long}, {Teague},
  {Booth}, {Walsh}, {Wilner}, {M{\'e}nard}, {Cataldi}, {Czekala}, {Bae},
  {Huang}, {Bergner}, {Ilee}, {Benisty}, {Le Gal}, {Loomis}, {Tsukagoshi},
  {Liu}, {Yamato}, \& {Aikawa}}]{Sierra2021}
{Sierra}, A., {P{\'e}rez}, L.~M., {Zhang}, K., {et~al.} 2021, \apjs, 257, 14,
  \dodoi{10.3847/1538-4365/ac1431}

\bibitem[{{Simon} {et~al.}(2022){Simon}, {Blum}, {Birnstiel}, \&
  {Nesvorn{\'y}}}]{Simon2022}
{Simon}, J.~B., {Blum}, J., {Birnstiel}, T., \& {Nesvorn{\'y}}, D. 2022, arXiv
  e-prints, arXiv:2212.04509, \dodoi{10.48550/arXiv.2212.04509}

\bibitem[{{Squire} \& {Hopkins}(2020)}]{Squire&Hopkins2020}
{Squire}, J., \& {Hopkins}, P.~F. 2020, \mnras, 498, 1239,
  \dodoi{10.1093/mnras/staa2311}

\bibitem[{{Stammler} {et~al.}(2019){Stammler}, {Dr{\k{a}}{\.z}kowska},
  {Birnstiel}, {Klahr}, {Dullemond}, \& {Andrews}}]{Stammler2019}
{Stammler}, S.~M., {Dr{\k{a}}{\.z}kowska}, J., {Birnstiel}, T., {et~al.} 2019,
  \apjl, 884, L5, \dodoi{10.3847/2041-8213/ab4423}

\bibitem[{{Suzuki} {et~al.}(2016){Suzuki}, {Ogihara}, {Morbidelli}, {Crida}, \&
  {Guillot}}]{Suzuki2016}
{Suzuki}, T.~K., {Ogihara}, M., {Morbidelli}, A., {Crida}, A., \& {Guillot}, T.
  2016, \aap, 596, A74, \dodoi{10.1051/0004-6361/201628955}

\bibitem[{{Tabone} {et~al.}(2022){Tabone}, {Rosotti}, {Cridland}, {Armitage},
  \& {Lodato}}]{Tabone2022}
{Tabone}, B., {Rosotti}, G.~P., {Cridland}, A.~J., {Armitage}, P.~J., \&
  {Lodato}, G. 2022, \mnras, 512, 2290, \dodoi{10.1093/mnras/stab3442}

\bibitem[{{Tazzari} {et~al.}(2021){Tazzari}, {Clarke}, {Testi}, {Williams},
  {Facchini}, {Manara}, {Natta}, \& {Rosotti}}]{Tazzari2021}
{Tazzari}, M., {Clarke}, C.~J., {Testi}, L., {et~al.} 2021, \mnras, 506, 2804,
  \dodoi{10.1093/mnras/stab1808}

\bibitem[{{Tazzari} {et~al.}(2016){Tazzari}, {Testi}, {Ercolano}, {Natta},
  {Isella}, {Chandler}, {P{\'e}rez}, {Andrews}, {Wilner}, {Ricci}, {Henning},
  {Linz}, {Kwon}, {Corder}, {Dullemond}, {Carpenter}, {Sargent}, {Mundy},
  {Storm}, {Calvet}, {Greaves}, {Lazio}, \& {Deller}}]{Tazzari2016}
{Tazzari}, M., {Testi}, L., {Ercolano}, B., {et~al.} 2016, \aap, 588, A53,
  \dodoi{10.1051/0004-6361/201527423}

\bibitem[{{Teague} {et~al.}(2018){Teague}, {Bae}, {Bergin}, {Birnstiel}, \&
  {Foreman-Mackey}}]{Teague2018}
{Teague}, R., {Bae}, J., {Bergin}, E.~A., {Birnstiel}, T., \& {Foreman-Mackey},
  D. 2018, \apjl, 860, L12, \dodoi{10.3847/2041-8213/aac6d7}

\bibitem[{{Teague} {et~al.}(2021){Teague}, {Bae}, {Aikawa}, {Andrews},
  {Bergin}, {Bergner}, {Boehler}, {Booth}, {Bosman}, {Cataldi}, {Czekala},
  {Guzm{\'a}n}, {Huang}, {Ilee}, {Law}, {Le Gal}, {Long}, {Loomis},
  {M{\'e}nard}, {{\"O}berg}, {P{\'e}rez}, {Schwarz}, {Sierra}, {Walsh},
  {Wilner}, {Yamato}, \& {Zhang}}]{Teague2021}
{Teague}, R., {Bae}, J., {Aikawa}, Y., {et~al.} 2021, \apjs, 257, 18,
  \dodoi{10.3847/1538-4365/ac1438}

\bibitem[{{Testi} {et~al.}(2014){Testi}, {Birnstiel}, {Ricci}, {Andrews},
  {Blum}, {Carpenter}, {Dominik}, {Isella}, {Natta}, {Williams}, \&
  {Wilner}}]{Testi2014}
{Testi}, L., {Birnstiel}, T., {Ricci}, L., {et~al.} 2014, in Protostars and
  Planets VI, ed. H.~{Beuther}, R.~S. {Klessen}, C.~P. {Dullemond}, \&
  T.~{Henning}, 339--361, \dodoi{10.2458/azu_uapress_9780816531240-ch015}

\bibitem[{{Trapman} {et~al.}(2022){Trapman}, {Zhang}, {van't Hoff},
  {Hogerheijde}, \& {Bergin}}]{Trapman2022}
{Trapman}, L., {Zhang}, K., {van't Hoff}, M. L.~R., {Hogerheijde}, M.~R., \&
  {Bergin}, E.~A. 2022, \apjl, 926, L2, \dodoi{10.3847/2041-8213/ac4f47}

\bibitem[{{Tychoniec} {et~al.}(2020){Tychoniec}, {Manara}, {Rosotti}, {van
  Dishoeck}, {Cridland}, {Hsieh}, {Murillo}, {Segura-Cox}, {van Terwisga}, \&
  {Tobin}}]{Tychoniec2020}
{Tychoniec}, {\L}., {Manara}, C.~F., {Rosotti}, G.~P., {et~al.} 2020, \aap,
  640, A19, \dodoi{10.1051/0004-6361/202037851}

\bibitem[{{Veronesi} {et~al.}(2021){Veronesi}, {Paneque-Carre{\~n}o}, {Lodato},
  {Testi}, {P{\'e}rez}, {Bertin}, \& {Hall}}]{Veronesi2021}
{Veronesi}, B., {Paneque-Carre{\~n}o}, T., {Lodato}, G., {et~al.} 2021, \apjl,
  914, L27, \dodoi{10.3847/2041-8213/abfe6a}

\bibitem[{{Vlemmings} {et~al.}(2019){Vlemmings}, {Lankhaar}, {Cazzoletti},
  {Ceccobello}, {Dall'Olio}, {van Dishoeck}, {Facchini}, {Humphreys},
  {Persson}, {Testi}, \& {Williams}}]{Vlemmings2019}
{Vlemmings}, W.~H.~T., {Lankhaar}, B., {Cazzoletti}, P., {et~al.} 2019, \aap,
  624, L7, \dodoi{10.1051/0004-6361/201935459}

\bibitem[{{Weidenschilling}(1977)}]{Weidenschilling1977}
{Weidenschilling}, S.~J. 1977, \mnras, 180, 57, \dodoi{10.1093/mnras/180.2.57}

\bibitem[{{Wichittanakom} {et~al.}(2020){Wichittanakom}, {Oudmaijer},
  {Fairlamb}, {Mendigut{\'\i}a}, {Vioque}, \& {Ababakr}}]{Wichittanakom2020}
{Wichittanakom}, C., {Oudmaijer}, R.~D., {Fairlamb}, J.~R., {et~al.} 2020,
  \mnras, 493, 234, \dodoi{10.1093/mnras/staa169}

\bibitem[{{Woitke} {et~al.}(2016){Woitke}, {Min}, {Pinte}, {Thi}, {Kamp},
  {Rab}, {Anthonioz}, {Antonellini}, {Baldovin-Saavedra}, {Carmona}, {Dominik},
  {Dionatos}, {Greaves}, {G{\"u}del}, {Ilee}, {Liebhart}, {M{\'e}nard},
  {Rigon}, {Waters}, {Aresu}, {Meijerink}, \& {Spaans}}]{Woitke2016}
{Woitke}, P., {Min}, M., {Pinte}, C., {et~al.} 2016, \aap, 586, A103,
  \dodoi{10.1051/0004-6361/201526538}

\bibitem[{{Xu} \& {Bai}(2022{\natexlab{a}})}]{Xu&Bai2022b}
{Xu}, Z., \& {Bai}, X.-N. 2022{\natexlab{a}}, \apjl, 937, L4,
  \dodoi{10.3847/2041-8213/ac8dff}

\bibitem[{{Xu} \& {Bai}(2022{\natexlab{b}})}]{Xu&Bai2022a}
---. 2022{\natexlab{b}}, \apj, 924, 3, \dodoi{10.3847/1538-4357/ac31a7}

\bibitem[{{Yang} {et~al.}(2017){Yang}, {Johansen}, \& {Carrera}}]{Yang2017}
{Yang}, C.-C., {Johansen}, A., \& {Carrera}, D. 2017, \aap, 606, A80,
  \dodoi{10.1051/0004-6361/201630106}

\bibitem[{{Yang} {et~al.}(2018){Yang}, {Mac Low}, \& {Johansen}}]{Yang2018}
{Yang}, C.-C., {Mac Low}, M.-M., \& {Johansen}, A. 2018, \apj, 868, 27,
  \dodoi{10.3847/1538-4357/aae7d4}

\bibitem[{{Youdin} \& {Johansen}(2007)}]{Youdin&Johansen2007}
{Youdin}, A., \& {Johansen}, A. 2007, \apj, 662, 613, \dodoi{10.1086/516729}

\bibitem[{{Youdin} \& {Goodman}(2005)}]{Youdin&Goodman2005}
{Youdin}, A.~N., \& {Goodman}, J. 2005, \apj, 620, 459, \dodoi{10.1086/426895}

\bibitem[{{Zhang} {et~al.}(2021){Zhang}, {Booth}, {Law}, {Bosman}, {Schwarz},
  {Bergin}, {{\"O}berg}, {Andrews}, {Guzm{\'a}n}, {Walsh}, {Qi}, {van't Hoff},
  {Long}, {Wilner}, {Huang}, {Czekala}, {Ilee}, {Cataldi}, {Bergner}, {Aikawa},
  {Teague}, {Bae}, {Loomis}, {Calahan}, {Alarc{\'o}n}, {M{\'e}nard}, {Le Gal},
  {Sierra}, {Yamato}, {Nomura}, {Tsukagoshi}, {P{\'e}rez}, {Trapman}, {Liu}, \&
  {Furuya}}]{Zhang2021}
{Zhang}, K., {Booth}, A.~S., {Law}, C.~J., {et~al.} 2021, \apjs, 257, 5,
  \dodoi{10.3847/1538-4365/ac1580}

\bibitem[{{Zhu} \& {Yang}(2021)}]{Zhu&Yang2021}
{Zhu}, Z., \& {Yang}, C.-C. 2021, \mnras, 501, 467,
  \dodoi{10.1093/mnras/staa3628}

\bibitem[{{Zsom} {et~al.}(2010){Zsom}, {Ormel}, {G{\"u}ttler}, {Blum}, \&
  {Dullemond}}]{Zsom2010}
{Zsom}, A., {Ormel}, C.~W., {G{\"u}ttler}, C., {Blum}, J., \& {Dullemond},
  C.~P. 2010, \aap, 513, A57, \dodoi{10.1051/0004-6361/200912976}

\bibitem[{{Zubko} {et~al.}(1996){Zubko}, {Mennella}, {Colangeli}, \&
  {Bussoletti}}]{Zubko1996}
{Zubko}, V.~G., {Mennella}, V., {Colangeli}, L., \& {Bussoletti}, E. 1996,
  \mnras, 282, 1321, \dodoi{10.1093/mnras/282.4.1321}

\end{thebibliography}
\bibliographystyle{aasjournal}



\end{document}